# Just-DNA-Seq, open-source personal genomics platform: longevity science for everyone.

Kulaga Anton[*, @, 1,2,3,4], Borysova Olga[*,4,6], Karmazin Alexey [4,7], Koval Maria[3,4], Usanov Nikolay[3,4], Fedorova Alina[3], Evfratov Sergey[3,], Pushkareva Malvina[3], Ryangguk Kim[8], Tacutu Robi [2, 5,]

1. Institute for Biostatistics and Informatics in Medicine and Ageing Research, Rostock University Medical Center, Rostock, Germany
2. Institute of Biochemistry of the Romanian Academy
3. International Longevity Alliance (ILA)
4. SecvADN SRL
5. CellFabrik SRL
6. MitoSpace
7. M. Glushkov Institute of Cybernetics of National Academy of Sciences of Ukraine
8. Oak Bioinformatics, LLC

\* The contribution of these authors is considered to be equal

@ - supervising and corresponding author

# Abstract

Genomic data has become increasingly accessible to the general public with the advent of companies offering whole genome sequencing at a relatively low cost. However, their reports are not verifiable due to a lack of crucial details and transparency: polygenic risk scores do not always mention all the polymorphisms involved. Simultaneously, tackling the manual investigation and interpretation of data proves challenging for individuals lacking a background in genetics. Currently, there is no open-source or commercial solution that provides comprehensive longevity reports surpassing a limited number of polymorphisms. Additionally, there are no ready-made, out-of-the-box solutions available that require minimal expertise to generate reports independently.

To address these issues, we have developed the Just-DNA-Seq open-source genomic platform. Just-DNA-Seq aims to provide a user-friendly solution to genome annotation by allowing users to upload their own VCF files and receive annotations of their genetic variants and polygenic risk scores related to longevity. We also created GeneticsGenie custom GPT that can answer genetics questions based on our modules. With the Just-DNA-Seq platform, we want to provide full information regarding the genetics of long life: disease-predisposing variants, that can reduce lifespan and manifest at different age (cardiovascular, oncological, neurodegenerative diseases, etc.), pro-longevity variants and longevity drug pharmacokinetics. In this research article, we will discuss the features and capabilities of Just-DNA-Seq, and how it can benefit individuals looking to understand and improve their health.

# Graphical abstract

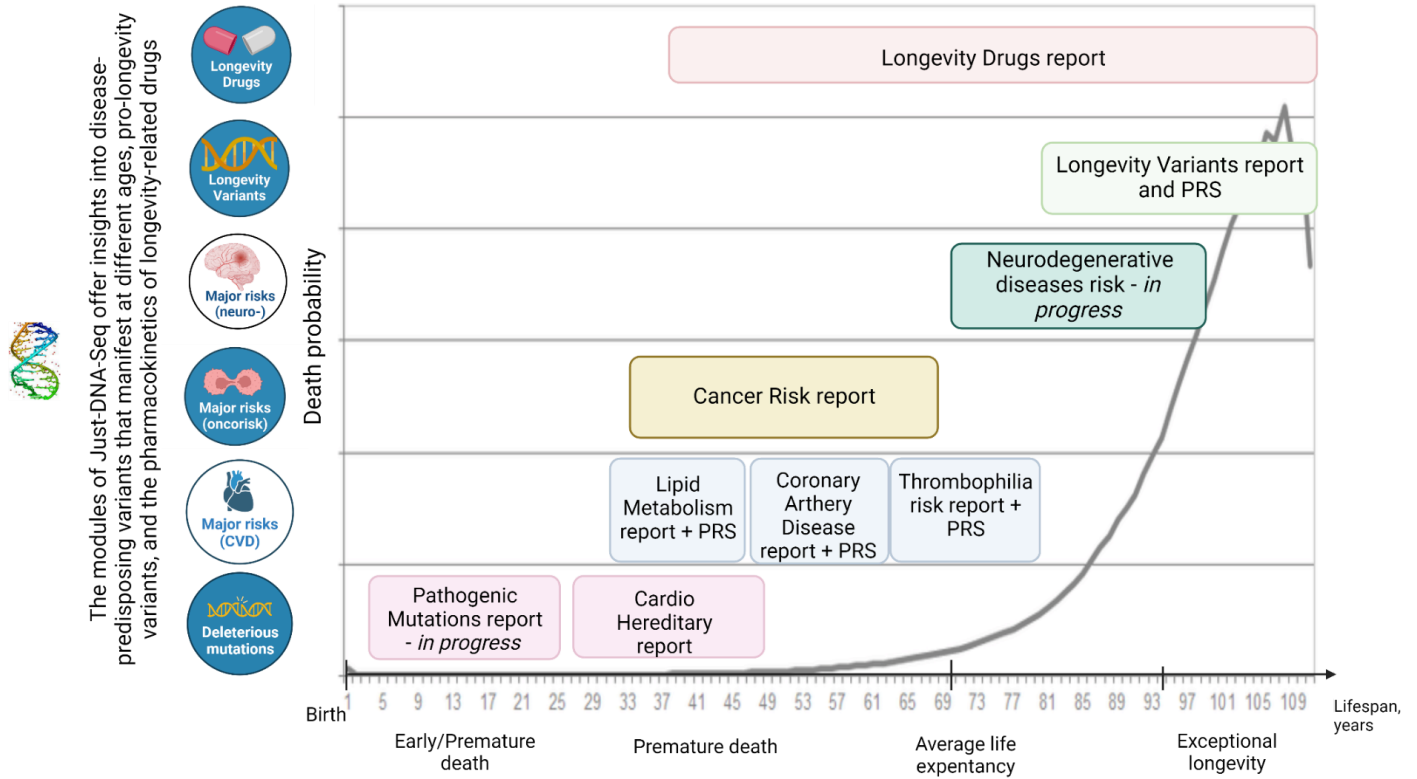

# Introduction

Personal genomics has the potential to revolutionize the field of medicine by providing individuals with personalized health information based on their unique genetic makeup. Bioinformatic and biological studies have produced significant scientific breakthroughs using the available wealth of genomic information. Since completing the Human Genome Project in 2003, genomic sequencing has become affordable and accessible. In addition, many companies offer direct-to-consumer genetic testing, which provides health, trait, and ancestry information to the public (Whitley et al., 2020).

Moreover, personal genomics offers a wealth of valuable insights into the genetic factors that influence longevity. This can lead to improved disease prevention and management strategies and a better understanding of individual responses to drugs and treatments. Understanding your longevity potential through personalized genomics becomes incredibly significant, enabling individuals to take proactive steps to manage their health, make informed lifestyle choices, and implement targeted preventive measures. By leveraging this knowledge, individuals can maximize their chances of living longer and healthier lives. However, it is important to note that predictions of our platform should be used for research and information purposes only. If you intend to adjust your treatment based on this information, consulting with a doctor or healthcare professional is strongly recommended.

The question of how much genetics contributes to longevity is a topic of inquiry. In terms of heritability, it is crucial to distinguish between lifespan, which refers to age at death, and longevity (Z. D. Zhang et al., 2020). Exceptional longevity phenomenon refers to individuals who live to 100 years and older, experiencing delayed onset of disability until their nineties, and for those surpassing 105 years, they also exhibit a progressive compression of morbidity, indicating a longer period of healthy living at extreme ages (Sebastiani & Perls, 2012). Family studies have long since shown that heritable factors contribute significantly to human longevity (Lunetta et al., 2007). A 2012 study suggests, among other things, that this complex phenotype may be dissected into several subphenotypes associated with exceptional longevity (Sebastiani et al., 2012). It was shown (van den Berg et al., 2019) that longevity is transmitted as a quantitative genetic trait among survivors up to the top 10% of their birth cohort, so the genetic component of lifespan is particularly strong (35%) in centenarians. Genome-wide association studies (GWAS) support a multi-factor model of longevity, in which longevity is influenced by many genetic variants (Pilling et al., 2016). Additionally, the impact of genes on longevity may involve trade-off-like effects on different health traits (Yashin et al., 2015). Some lower-effect genetic factors may interact with each other, giving statistically significant cumulative effect (Yashin et al., 2010), adding an additional layer of complexity to correlating genetics with longevity.

Despite those problems, in recent years, some studies have explored the correlation of specific alleles of the human genome with longevity. A meta-analysis of several independent GWAS published in 2015 has shown that genes FOXO3 and APOE are strong candidates for being associated with longevity (Broer et al., 2014). Two new SNPs associated with longevity were discovered in a 2016 GWAS on the Han Chinese population (Zeng et al., 2016). A GWAS conducted in 2017 on many biobank participants from the UK identified 25 genome loci with a statistically significant association with longevity (Pilling et al., 2017). Review and meta-analysis have shown that five polymorphisms, ACE rs4340, APOE ε2/3/4 rs7412, rs429358, FOXO3A rs2802292, KLOTHO KL-VS rs9536314, and IL6 rs1800795 were significantly associated with exceptional longevity (Revelas et al., 2018). But Caruso et al., in their review study (2022), mentioned that only two genes, APOE and FOXO3A, involved in the protection of cardiovascular diseases, have been shown to be associated with longevity in nearly all studies. Except for common SNPs, the genetic component of extreme human longevity constitutes rare coding variants in pathways that protect against ageing, which were demonstrated by whole-exome sequences of a large cohort of Ashkenazi Jewish centenarians (Lin et al., 2021). Recently, high-throughput sequencing analysis also identified four novel variants in nuclear-encoded mitochondrial genes associated with longevity: MTOR (Y2396Lfs29), CPS1 (T1406N), MFN2 (G548), and LRPPRC (S1378G) (Gonzalez, 2023).

As a result, the data on longevity genetics is rapidly expanding, serving as the foundation for creating comprehensive databases collected in Human Ageing Genomic Resources (HAGR) (Tacutu et al., 2017). However, those databases were designed primarily for science and research purposes and lacked capabilities for automated annotation of personal genomes. The main concerns regarding the commercial solutions are the need for more transparency and privacy of the results (Bonomi et al., 2020). Users cannot fully trace how exactly their genetic risks were computed, they do not know if their genetic data will be sold elsewhere, and their abilities to explore their genomes beyond already provided reports are very limited. Another ethical concern arises as individuals may confront information that goes beyond their ability to manage, including the possibility of inaccuracies. This underscores the importance of discussing genetic findings with a healthcare professional, emphasizing the necessity for clear communication and support in navigating the complexities of this personal data. Transparency of our platform eases such communication.

To address the issues mentioned above and to enable both ageing researchers and longevity enthusiasts to apply longevity genetics to personal genomes, we developed Just-DNA-Seq - an open-source personal genomics platform with a focus on longevity.

# Results

The main goal of the Just-DNA-Seq platform is to provide full information regarding the genetics of long life. Based on the hypothesis of the genetic basis of exceptional longevity, proposed by Sebastiani & Perls (2012), increasing age is accompanied by a decrease of disease-predisposing variants and by a selection for longevity-associated variants that can counter the deleterious effects of genetic and environmental factors) and an increase of longevity-enhancing variants that afford protection against basic mechanisms of ageing, slow the rate of ageing, and delay the onset of age-related diseases and syndromes.

We think that in terms of longevity, it is essential to analyze:

1. **Disease-predisposing ("anti-longevity") gene variants and PRSs.**
   The main age-related diseases that cause death include cardiovascular diseases, cancer, and neurodegenerative diseases. They all tend to manifest at a particular age, leading to increased mortality. Genetic testing can be instrumental in prevention as it allows for identifying individuals who may have an increased genetic risk for certain diseases. This early identification enables personalized preventive measures and interventions to be implemented, such as lifestyle modifications, regular screening, and targeted therapies.
2. **Longevity-enhancing ("pro-longevity") gene variants and PRSs**.
   In this category, we include variants that are thought to be protective from various diseases and are involved in the basic mechanisms of ageing. These variants are gathered from various longevity centenarian studies and databases. In this section, we also analyze SNPs and Polygenic risk scores.
3. **Pharmacogenetics of drugs.**
   There are some drugs, like metformin and rapamycin, that are thought to be potentially pro-longevity. However, there can be personalized risks and a need for dosage correction according to genotype. Testing pharmacogenetics becomes essential to understand individual metabolism and identify personalized risks, optimizing drug treatments and promoting better health outcomes.

To address this, we gathered this information in the Just-DNA-Seq platform. Just-DNA-Seq utilizes genetic polymorphisms sourced from various databases and supplements them with new data extracted from recent articles on PubMed, employing search keywords such as "genetics human longevity," "longevity predisposing variants," and "lifespan human genetics." In particular, it substantially expands the LongevityMap database by adding details from meta-analyses published after the year 2017 and prioritizing gene variants based on their respective 'weights.' Just-DNA-Seq provides information about the longevity-associated gene variants, variants predisposing to chronic age-related diseases, Polygenic Risk Scores of Longevity (Tesi et al., 2020) chronic age-related diseases, and personal responses to longevity-related drugs. This data is represented as separate reports. We also created a "Genetics Genie" custom GPT which is available in the ChatGPT store and answers genetic questions based on Just-DNA-Seq modules.

To sum up, for today, Just-DNA-Seq is the first and only genetic platform providing user-friendly information regarding different aspects of longevity.

## Just-DNA-Seq functionality and structure

Just-DNA-Seq is structured as a modular, web-based platform that allows users to easily upload their data and visualize and annotate genomic variations with a focus on longevity. To work with Just-DNA-Seq, a person can upload a VCF file of variants and choose from a range of Just-DNA-Seq reports.

The platform consists of a variety of functional modules that can be downloaded as reports:

1. Longevity variants module - shows a list of longevity-related SNPs (single nucleotide polymorphisms).
2. PRS (polygenic risk score) module - shows PRS for different diseases and longevity.
3. Longevity Drugs module - shows a list of snips that influence drug dosage.
4. Major Health risks module, subdivided into thrombophilia risk module, inherited cardiovascular diseases module, coronary artery disease module, germline cancer risk module, and lipid metabolism module.

These modules can be used in combination or separately, providing a flexible and scalable solution for functional annotation and variant prioritization. Certain risks characterize each life period that can affect lifespan, so getting genetic information regarding the risks and their management is important to obtain longevity.

Longevity Variants Module

Functionality and interface

The Longevity Variants module uses curated and rearranged data from the Longevity Map database and additional information from other sources. The module also relies on the ClinGene, dbSNP, and ClinVar modules for gene-based annotation. The longevity variants report (Figure 3) consists of a brief description, a list of gene variants in an annotated genome, and very general recommendations about additional tests that can be made.

As of today, the LongevityMap database of human genetic association studies of longevity contains about 3144 longevity-associated gene variants in 884 genes (Tacutu et al., 2017), and the more recent sources mention 113 genome-wide significant longevity and 14 529 age-related disease variants (Brooks-Wilson, 2013).

After curation, the data from the Longevity Map database is presented as a list of SNPs that are statistically significant with regard to human longevity. The table includes information on a person's genotype for each SNP, the population for which the significance of the variant was shown, the gene where the SNP is located, and its zygosity. Each variant is prioritized in the list based on its weight. A positive weight is marked in green to indicate a pro-longevity variant, while a negative weight is marked in red to indicate a variant associated with decreased lifespan. The weight value is an integrative parameter that considers the number of studies that have shown the significance of the variant, its p-value for different studies, and the number of populations for which the variant is significant. In the list of Longevity variants, the SNPs are arranged according to their relative weight value.

![Longevity Variants report interface]

| | RSID | Population | Gene | Your genotype | Ref allele | Alt allele | Zygosity | Weight |
|---|---|---|---|---|---|---|---|---|
| + | rs2267723 | Danish | GHRHR | G/A | A | G | het | -0.11 |
| + | rs5771675 | American (Caucasian) | FAM19A5 | G/A | A | G | het | 0.065 |

**Figure 3. Longevity Variants report interface**. The report contains a general description of the Longevity variants module, Longevity Polygenic Risk scores (PRS), and Longevity variants categorized into 12 distinct pathways. The general description of the module and the Longevity PRS can be accessed by selecting the respective accordion. The PRS encompasses an analysis of 332 variants that significantly discriminated between centenarians and older adults. Each Longevity pathway features "Description", "Results", and "Recommendations" sections. The "Description" section provides information about the pathway's role in Longevity. The "Results" section contains a table with a list of SNPs (Single Nucleotide Polymorphisms) associated with exceptional longevity, including the population in which this association was demonstrated, the user's genotype for this SNP, and the relative weight of this polymorphism. The relative weight is an integrated parameter that considers the number of studies on this polymorphism in relation to its association with longevity, p-values, and the number of populations for which the association is shown. Polymorphisms favorable for longevity are marked in green, while unfavorable ones are marked in red. The "Recommendations" section contains general recommendations that can vary depending on individual SNPs.

## Data curation and implementation details

The process of creating the module involved filtering significant variants from the Longevity Map database, rearranging data, and adding information on effect alleles regarding longevity traits. In particular, we edited 876 records, added 4 additional fields for each record, and provided 50 entries on top of existing ones.

This information was combined with recent publications' data to create a more complete and up-to-date resource.

After filtering the significant gene variants from the Longevity Map database, for each identification number (ID) corresponding to a specific single nucleotide polymorphism (SNP), we included additional fields containing information about the reference and alternative alleles, as well as all possible genotypes for that SNP. We then manually curated the data to identify alleles or genotypes associated with longer or shorter lifespans.

To assign a value representing the impact on longevity, we assigned a "longevity genotype weight" to each genotype for a particular SNP. A value of 0 represented a neutral genotype, 0.5 indicated a heterozygous genotype with a pro-longevity allele, 1 represented a homozygous genotype with two pro-longevity alleles, -0.5 indicated a heterozygous genotype with an anti-longevity allele, and -1 represented a homozygous genotype with two anti-longevity alleles. If the association between genotype and longevity was only observed for heterozygous variants, the value assigned was solely for the heterozygous genotype, while all other genotypes received a value of 0.

All the records got "SNP weight", which is an integrative parameter that considers the number of studies that have shown the significance of the variant, its p-value for different studies, and the number of populations for which the variant is significant. The weight value the person sees in a Report for the genotype of this particular SNP is a product of the multiplication of "SNP weight" to "longevity genotype weight".

## Polygenic Risk Scores Module

### Functionality and interface

A polygenic risk score (PRS), also known as a polygenic score (PGS) or genetic risk score (GRS), is an assessment of a person's genetic risk for a particular trait or disease. It is derived by combining and quantifying the impact of numerous common variants associated with the condition in the genome, each of which can have a small impact on an individual's genetic risk for a particular disease or condition. The weighted sum of a group of genetic variants is how a PRS is commonly created. Higher scores indicate higher risk, and the final score is generally normally distributed in the overall population (Collister et al., 2022).

PRS (polygenic risk score) module uses information from the Polygenic Score (PGS) Catalog (Lambert et al., 2021) and represents it as a percentile score, which reflects an individual's risk relative to other individuals in the population. The PRS module for estimating longevity and disease risks is based on analyzing multiple genetic variants associated with ageing, longevity, and ageing-associated diseases. By integrating this information, the module provides a comprehensive assessment of an individual's risk for a range of diseases and conditions, including age-related diseases such as cardiovascular disease, inflammation risk, and cancer.

Percentiles are calculated in several steps. In the first step, all the weights were extracted, along with their probabilities, from the gnomad database. In the second step, the population was sampled. For each weight, two probability checks were made. If both pass, it is equivalent to homozygote, if one passes to heterozygote, and if none passes to the absence of the effect allele. All the weights summed respectively 2, 1, or zero times. The resulting number corresponds to a single sample. This process was repeated 10000 times to get a big enough distribution, and after that, a statistical library (SciPy) was applied to get mean and sigma, assuming it is a normal distribution. Then, percentiles were calculated as percent point function.

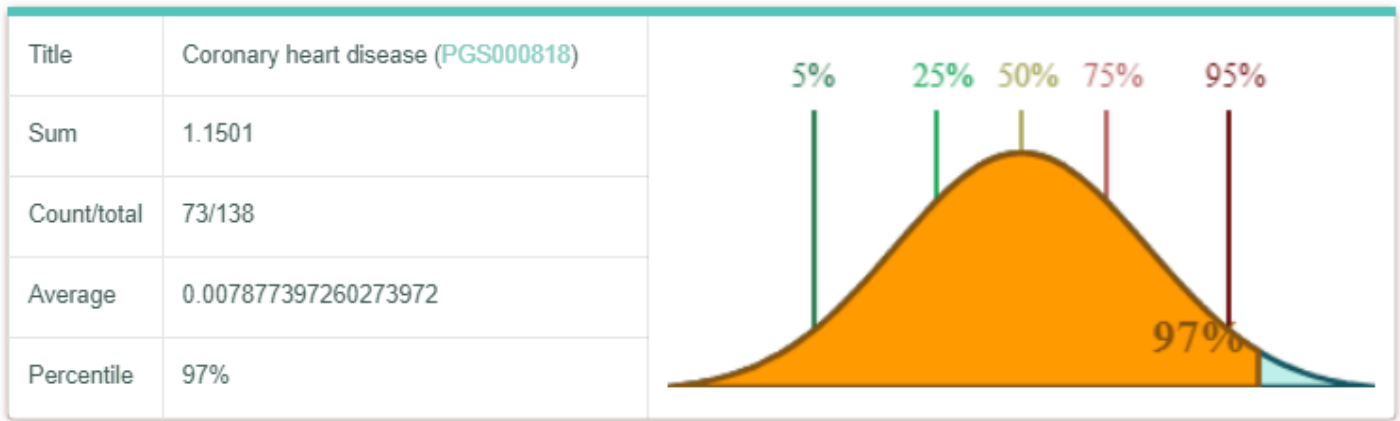

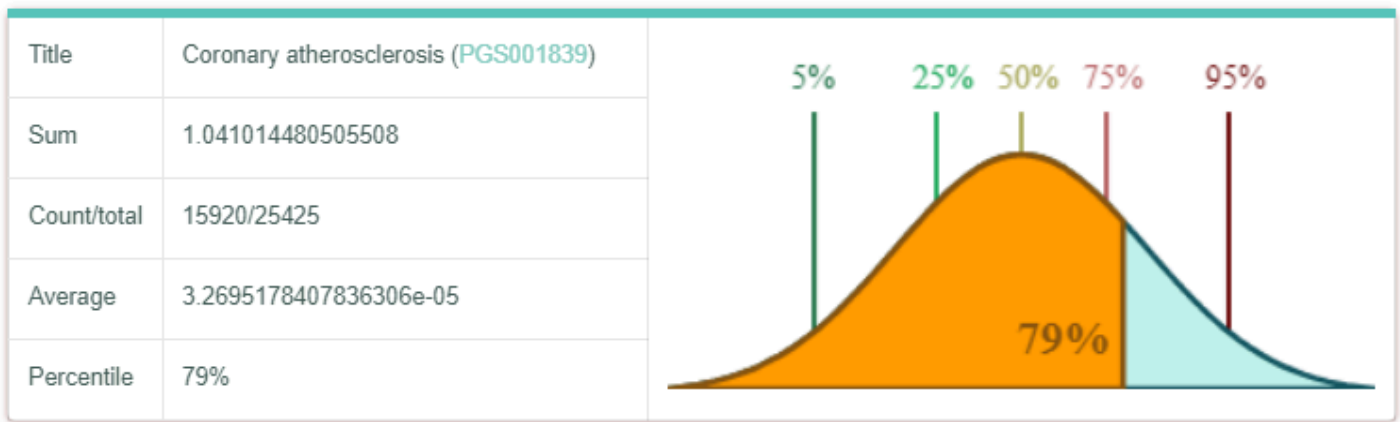

**Figure 4**. The PRS (Polygenic Risk Score) report module provides information for each calculated PRS, represented as a percentile within a given population. This information includes the feature's name (disease), its number in the PGS catalog, the score sum, the number of gene variants detected in the genome out of the total included in the score, the average value, and the percentile in which the result for that specific genotype falls within the population. The sum is the addition of all the weights of PRS that appear in the genome. In the case of a homogeneous allele, it is multiplied by 2. Average is the sum divided by count * 2.

Data curation and implementation details

The data curation process for the Polygenic risk scores module involves collection, processing, and control stages. Relevant PRSs are extracted from the open database of polygenic scores, their consistency with related genome-wide association studies is checked, and metadata is collected. Genome build adjustments and conversions are performed if required.

The module uses tools from the Polygenic Score Catalog Calculator (pgsc_calc) (Wingfield et al.) to define the most similar population by handling the 1000 Genomes resource (Auton et al.) as a population reference panel and obtaining genetic ancestry data using PCA.

Longevity Drugs Module

Functionality and interface

The Longevity Drug module utilizes information from two key sources: the PharmGKB database (Whirl‑Carrillo et al., 2021) and ClinVar (Landrum et al., 2017) drug information. The PharmGKB database provides information on the genetic basis of drug response, including information on genes and variants that are known to influence drug response. The ClinVar drug information provides information on the clinical significance of genetic variants, including information on drug response and the potential impact of genetic variations on drug efficacy and safety.

This information is also integrated into the Longevity Drug module. The combination of information from the PharmGKB database and ClinVar drug information provides users of the Longevity Drug module with a comprehensive resource for information on the genetic basis of drug response, possible complications, and the need of drug dose adjustment.

**Figure 5. The Longevity Drugs report** provides information about pharmaceutical drugs for which the response may be altered in individuals with specific genetic variants. The information is organised into columns, including the specific rsID, the name of the medication associated with this genetic variant, and the Phenotype category, indicating the specific changes linked to this variant, such as a decrease or increase in drug effectiveness or the presence of side effects associated with the medication. Additionally, a brief description of how the genetic variant affects the medication is provided.

### Data curation and implementation details

The curated data includes information on genes and variants that are known to influence drug response, as well as information on the clinical significance of genetic variants and the potential impact of genetic variations on drug efficacy and safety.

PharmGKB database contains the quantitative pharmacogenetic data in the Variant annotation summary section, table "study_parameters.tsv" as an indexed list of studies with effect size values (OR/RR/HR) and associated statistical parameters, such as p-values and group sizes. Each study has a unique ID "Variant Annotation ID" for cross-references between other PharmGKB tables. The second essential table is "var_drug_ann.tsv" in the same location.This table contains information about associations in which the variant affects a drug dose, response, metabolism, etc., and each variant has an index "Variant Annotation ID" – by this index, we can link these two tables one-to-one and get known measurable effects of genetic variants. After filtration, our tool performs trivial substitution of alleles from input and aggregate effects from all variants (with known variant-drug effect) for each drug; currently, effects from different variants for the same drug are just multiplied.

For analysis, we take input as table RSID:Allele/Allele, substitute these input alleles to the clear effect-variant-drug associations table, and produce substituted effect-variant-drug associations table. After the generation of the substituted table, we perform pivoting by each drug for integration of effects from different variants by simple multiplication. Also, we separate these computations between different kinds of Phenotype Categories (i.e. all Odd Ratios and Relative Risks are multiplied separately also with separation between

Metabolism/Pharmacokinetics and drug Efficacy). By this simple approach we provide an integrated numerical estimation of genetic influence on the drug metabolism and efficacy.

## Health Risks Modules

### Functionality and interface

Health risks modules are based on filter-based post-aggregators or table-based ones.

Germline cancer risk and hereditary cardiovascular diseases are filter-based modules. In particular, they are based on OpenCravat annotators with our custom filters that separate only essential and relevant information depending on different fields. The very first filter in both modules uses a list of genes provided as a .txt file in the data folder of the module (with 429 genes for the cancer risk module and 334 genes for the hereditary cardiovascular diseases module) (see Supplementary).

The next filter is based on the significance of an SNP using the results of ClinVar annotation. In the hereditary cardiovascular diseases module, we filter out only pathogenic or likely pathogenic SNPs, which are then displayed in a report. Meanwhile, in the germline cancer risk module, we filter genes with different significance, such as risk factors, protective SNPs, pathogenic factors as well as SNPs with conflicting interpretations of pathogenicity.

After that, for the hereditary cardiovascular diseases module, we filter out only damaging variants (based on SIFT prediction) related to arrhythmias and cardiomyopathy SNPs (based on Cardioboost-based annotator), and for the germline cancer risk module providing results are filtered out just based on ClinVar annotator.

The following modules are table-based: thrombophilia, lipid metabolism, and coronary arterial disease. Information retrieval is organized as a SQL query that connects a VCF record with table data by rsID. For this purpose, we use dbSNP annotator.

### Coronary artery Disease

> Description

> PRS

˅ Results

| | # | RSID | Gene | Risk Allele | Genotype | Pubmed ID | Population | P-Value | Weight |
|---|---|---|---|---|---|---|---|---|---|
| + | 1 | rs8055236 | CDH13 | G | G/G | PMID 17554300; PMID 19956433; PMID 20017983; | European | [PMID 17554300]: 6 x 10-6 | -1.7 |
| + | 2 | rs4977574 | CDKN2B-AS1 | G | G/A | PMID 17478681; PMID 21378990; PMID 21378988; PMID 24916648; PMID: 30278588 | European; Asian | [PMID 29263402]: 1 x 10-7; [PMID 21239051]: 1 x 10-14; | -1.3 |
| + | 3 | rs599839 | SORT1 | A | A/A | PMID: 32858814; PMID 33321069; PMID: 17634449; PMID 18262040; PMID 19380133; PMID 19660754; PMID 19750184; | European | [PMID 33321069]: 2 x 10-7 | -1.2 |
| + | 4 | rs17465637 | MIA3 | C | C/C | PMID: 35768776; PMID 19956433; PMID 21804106; PMID 21984477; PMID 21264445; | European; Iranian; | [PMID 17634449]: 1 x 10-6; [PMID 21378990]: 1 x 10-8; | -1.2 |
| + | 5 | rs1122608 | LDLR | G | G/G | PMID: 27664493; PMID 23380588; PMID 23202125; PMID 20810930; PMID 19956433; PMID: 33321069; PMID 21378990; | European; Asian | [PMID: 33321069]: 5 x 10-7; [PMID 21378990]: 1 x 10-9 | -1.0 |
| + | 6 | rs7250581 | | G | G/G | PMID 19955471; PMID 19956433; PMID 17554300; | European | | -0.8 |

**Figure 6**. **The Major Health Risk report** consists of several individual sub-reports: germline cancer risk, thrombophilia risk, coronary artery disease risk, lipid metabolism disorder risk, and hereditary heart disease risk. The report includes a list of polymorphisms and genes associated with an increased or decreased risk of specific diseases. It provides references to studies, the population for which the association was established, p-values, and the relative "weight" of each identified variant. Genetic variants that increase the risk of disease are highlighted in red, while those that decrease the risk are highlighted in green.

### Data curation and implementation details

Table-based modules are based on analyzing SNP data associated with the risk of developing lipid metabolism disorders, thrombophilia, and coronary artery diseases. These modules are manually curated to include the most significant SNP data related to these conditions. The inclusion criteria for selected SNPs were p-values from meta-studies, the number of studies for different populations. We also added information on risk and protective alleles, descriptions for gene functions, and possible health effects based on PubMed, SNPedia (Cariaso & Lennon, 2011), and GeneCards (Stelzer et al., 2016) data. The variants are prioritized based on their impact. In addition to being a manually-curated database, it includes data regarding effects on gene substitution from sources such as ClinVar, PubMed, and SNPedia

### Neural networks models

Deep neural networks (DNNs) have become increasingly popular in genomic and epigenomic research due to their ability to effectively use large-scale bioinformatics data and achieve highly accurate predictions and classifications. Nevertheless, DNN-based methods frequently encounter challenges due to their potential to explain the relationships between initial information in the first layer and final prediction (Talukder et al., 2020). The Just-DNA-seq ML module harnesses the potency of interpretable deep learning to elucidate phenotypes from genetic data. For this purpose, trained networks from GenNet (van Hilten et al., 2021), an open-source deep-learning framework for predicting phenotypes from genetic variants, are used.

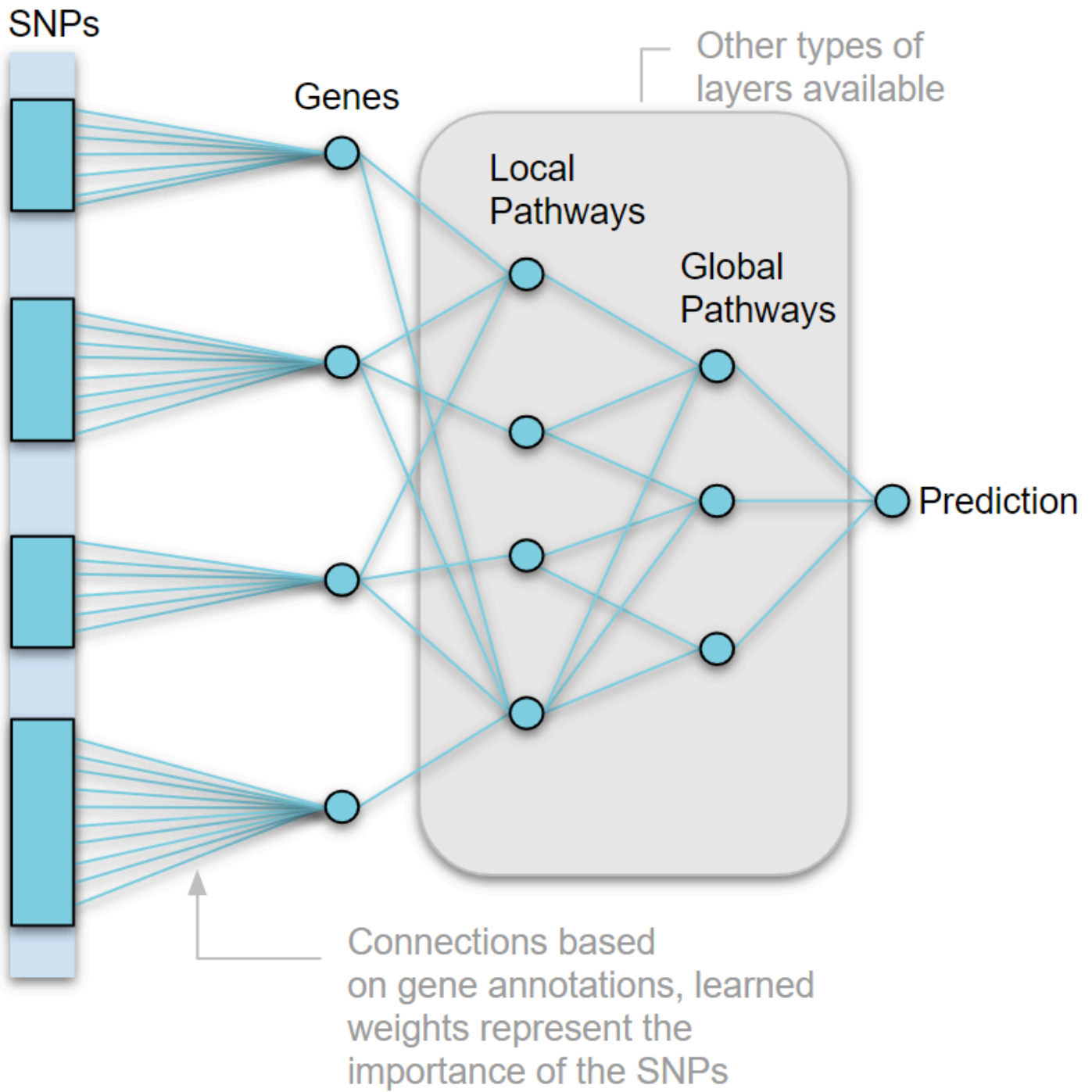

**Figure 7. GenNet framework architecture.** Prior biological knowledge to define connections between layers is utilised to create interpretable networks (i.e., SNPs are connected to their corresponding genes by using gene annotations, and genes are connected to their corresponding pathway by using pathway annotations)

System implementation and performance

Trained networks with higher performance were indicated for schizophrenia with GTEx layers knowledge and dementia with RefGene layers knowledge. The validation accuracy score for the schizophrenia network is 0.65 and AUC in the test set is 0.66 while the estimated theoretical upper limit for classification accuracy is between 0.73 for the schizophrenia dataset used for network training. The validation accuracy score for dementia is 0.58 and AUC in the test is 0.61 while the estimated theoretical upper limit for classification accuracy is 0.83.

Data curation and implementation details

The weights for best-predictive networks were obtained from publicly available GenNet repositories. To get data, subjected to prediction and fitted to the first layer of a network, the initial VCF file is processed through imputation (Browning et al., 2018), filtering, and conversion.

## General Platform Design

The Just-DNA-Seq platform comprises a set of open-source OakVar modules, bioinformatics pipelines, and additional source code libraries with documentation and tutorials. The core of Just-DNA-Seq is a set of modules on top of the OakVar genomic variant analysis platform (rkimoakbioinformatics, n.d.) (Fig 1). OakVar is a Python-based genomic variant analysis platform. It is a fork of OpenCRAVAT and is an open-source software tool for integrating, annotating, and analysing genomic data, focusing on identifying genetic variants associated with disease and other phenotypes. It has a modular architecture. Such implementation enables smooth integration with hundreds of other genomic modules and allows applying generic OakVar interfaces and filters to explore its results manually.

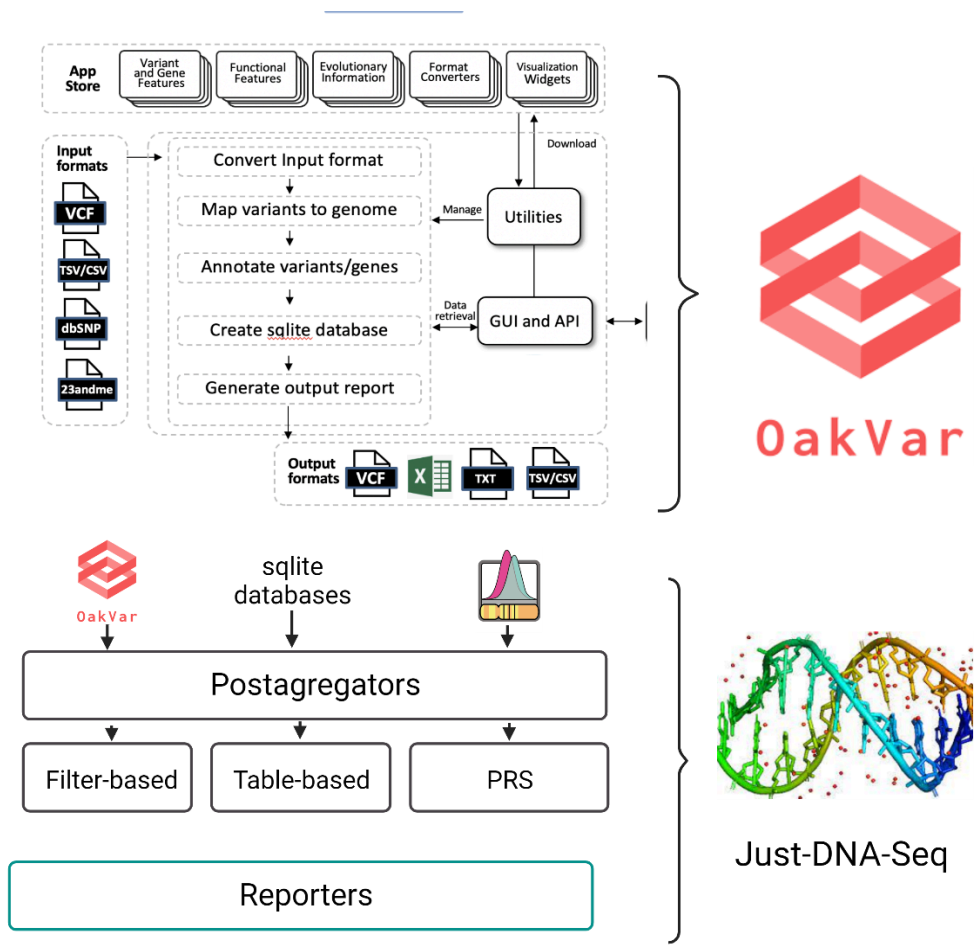

**Figure 1. Just-DNA-Seq architecture.** Just-DNA-Seq platform uses Open Cravat module naming rules and design patterns in order to create a report pipeline. We use post-aggregators because it is the lowest level where we have all the data we need. Each post-aggregator is responsible for logically separated parts of the report. At this stage, we do all intensive processing and save results in the SQLite database. In the reporter stage, we just convert the prepared data in the report with the help of Mako templater which is a lightweight and fast templating engine for Python. It makes the report generation process almost instantaneous when all the information is already precalculated.

Just-DNA-Seq can be run both in web-based and console modes. A typical workflow of personal genomic analysis (Figure 2) involves uploading user VCF files either provided by the sequencing providers or produced by our or other bioinformatic pipelines from raw fastq/bam files. Major steps of input format conversion, mapping, and annotations are conducted by the OakVar framework, while Just-DNA-Seq modules are focused on post-aggregation, PRS computation, and report generation.

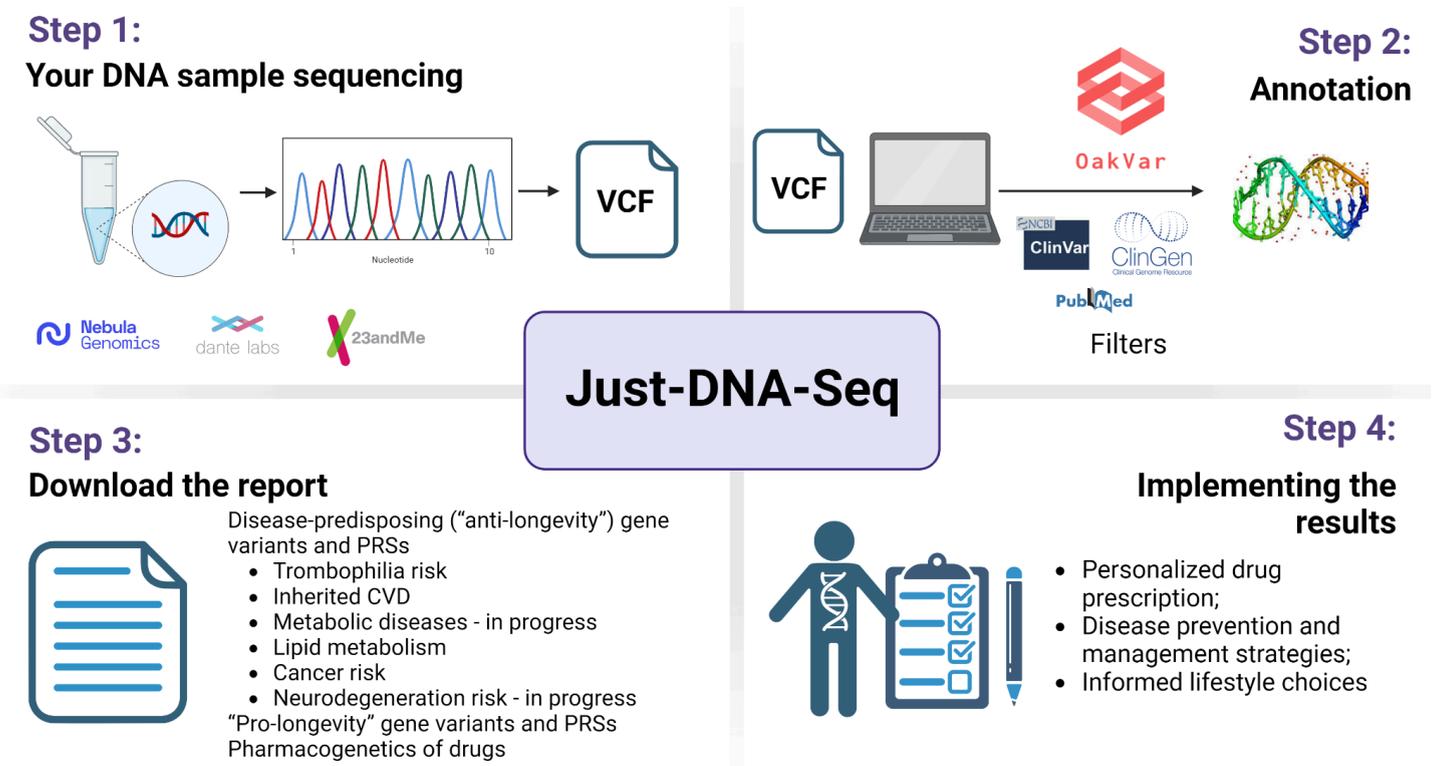

**Figure 2. Workflow of getting Just-DNA-Seq genome report.**
Users start by sequencing their genome with a commercially available service. Then the user can upload the resulting VCF file to the created OakVar account. After processing, the user can easily access a report containing information about gene variants and PRSs related to longevity.

Post-aggregators are divided into filter-based and table based. Filter-based post-aggregators are based on OakVar annotators with custom filters that separate only important and relevant information. It allows the integration of various genomic and functional databases, allowing users to easily annotate and visualise the impact of genomic variations on longevity. Just-DNA-Seq modules depend on dbSNP (RSID resolution), ClinVar (clinical significance and disease information), OMIM (mendelian inheritance), NCBI gene (general genetic annotation, PubMed (articles information), GnomAD (frequencies for variants, used for computing PRS scores), VCFInfo and other OakVar modules (complete list of dependencies of each module is provided in their dependencies described in yaml files). Just-DNA-Seq also provides various tools for prioritising variants based on their functional and clinical significance; it integrates functional impact prediction from SIFT, PolyPhen, and CADD scores, as well as variant frequency information from GnomAD for different populations, including both rare and common variants. The filter-based post-aggregators also consider the significance and importance of variants based on ClinVar annotation and OMIM information.

Table-based post-aggregators use curated databases. They compile data on Single Nucleotide Polymorphisms (SNPs) linked to a particular condition or disease by sourcing information from a variety of databases such as the GWAS catalog, ClinVar database, OMIM database, and conducting PubMed searches with pertinent keywords related to the specified condition or disease. The gathered information includes fields such as rsID, gene name, reference and alternative alleles, conclusions about the impact of the specific SNP on health

parameters, population details, PubMed links, and p-values. Information retrieval is organized as a SQL query that connects the vcf record with table data by rsID field. For this purpose, we use dbSNP annotator.

We employ manual curation, which involves staying updated on newly published data and database modifications. Additionally, we integrate information from recent research articles pertaining to the risk of developing a specific disease. We also incorporate gene-based information that has already been computed by other OakVar modules. This approach helps us to identify numerous SNPs that are associated with the risks of disease development. Most of the SNPs are present with their SNP IDs, dbSNP database, and the corresponding Oakvar Module is used for SNP resolution from genomic coordinates.

Architecturally, the modules include post-aggregator modules to annotate and summarise genetic variants, compute scores, and reporter modules to present results in a user-friendly format, including tables, graphs, and interactive visualizations.

We also provide PRS scores. PRS, short for polygenic risk score, represents a score that conveys information about an individual's genetic risk for a particular trait or disease, as determined by the cumulative effect of multiple genetic variants. The PRS scores used in our annotations are sourced from the PGS catalog (https://www.pgscatalog.org/) and are calculated from the genotype data and variant weights that are based on established genome-wide association studies.

PRS post-aggregator is a bit different from the two kinds of post-aggregators. We used neither filters with existing modules nor curated databases. It is similar to the map-reduce algorithm where we look for matching VCF records with PRS parameters in the first stage and then make a summation, calculating the average and finding the percentile in the second stage.

# Discussion

The genetics of longevity is intricate and still need to be fully comprehended. The duration of one's lifespan is affected by a multitude of genes, each of which has a limited impact, as well as environmental and lifestyle elements. A few genetic variations have been linked to an extended lifespan (Donlon et al., 2017), whereas others have been linked to a shorter lifespan. The heritability of an individual's lifespan and exceptional longevity is not the same, and the ability to live exceptionally long has a stronger genetic component (Z. D. Zhang et al., 2020). Lifespan can be affected by certain genetic variations that impact the likelihood of age-related illnesses (Bin-Jumah et al., 2022; Passarino et al., 2016).

Moreover, gene polymorphisms, or genetic variations, can affect how individuals respond to drugs, impacting longevity by reducing the drug efficacy or increasing the risk of adverse effects. Additionally, some drugs may have secondary effects on ageing and longevity, such as reducing inflammation or oxidative stress, which can also play a role in determining lifespan. There are several potential geroprotectors that can influence ageing by various pathways (Moskalev et al., 2022). Therefore, understanding how gene polymorphisms impact drug response is crucial for optimising treatment outcomes and potentially extending lifespan.

So the objective of Just-DNA-Seq is to create a comprehensive personal genomics platform that provides a complete picture from a current stage of our knowledge of how genes can affect an individual's longevity and lifespan.

The platform includes data on genetic variants that could potentially reduce lifespan (such as those increasing the risk of cardiovascular diseases, neurodegenerative diseases, and cancer), variants that may contribute to a longer lifespan and exceptional longevity, genes involved in drug metabolism, and polygenic risk scores (PRSs). It's important to note that this information will benefit individuals of any age. Genetics determine the risks of diseases, the likelihood of their manifestation varying with age, and the effectiveness of risk prevention strategies involves several factors, primarily relying on accurate awareness of their presence.

# Disease-predisposing gene variants and PRSs

Over the past century, the average human lifespan has extended, primarily attributed to reducing numerous infectious illnesses. Presently, more individuals succumb to age-related diseases than before (Juckett, 2009). Age-related diseases are conditions that become more common or severe as a person gets older. Some of the main age-related diseases include cardiovascular diseases, cancer, and neurodegenerative diseases. They all tend to manifest at a particular age, leading to increased mortality. A shift in focus towards the preventive approach was made in the last decade (Ray et al., 2022). Genetic testing can be instrumental in early preventive approaches as it allows for identifying individuals with an increased genetic risk for certain diseases. This early identification enables personalised preventive measures and interventions to be implemented, such as lifestyle modifications, regular screening, and targeted therapies.

Even so, the relationships between human longevity and genetic risks of complex diseases were not broadly studied, polygenic risk scores (PRSs) for 225 complex diseases/traits in a cohort with 2178 centenarians and 2299 middle-aged individuals showed the lower genetic risks of stroke and hypotension in centenarians (Hu et al., 2022).

To address the question regarding age-related diseases risks, Just-DNA-Seq has several modules that provide information regarding health risks. Reports that are already available: Cardio Hereditary Report, Lipid Metabolism, Thrombophilia, Coronary Artery Disease, and Germline Cancer Risk. This section also provides PRSs: Coronary atherosclerosis (PGS001839), Coronary heart disease (PGS000818), Blood clots or deep vein thrombosis (PGS000931).

Except for age-related diseases, which had heritability between 40-80 % (Melzer et al., 2019), we found it helpful to add in this section reports, providing information about inherited genetic conditions (heart diseases and some metabolic diseases). They are pretty rare but can cause sudden death if they remain undiagnosed and untreated.

Cardio Hereditary Report. Timely detection and diagnosis of heart disorders can lead to enhanced treatment options, help to prevent sudden cardiac death, and improve prognosis. This report analyses the most relevant genes, abnormalities of which may cause the following syndromes: Long and short QT, Brugada syndrome, catecholaminergic polymorphic ventricular tachycardia, cardiomyopathies dilated and hypertrophic, and congenital heart defects. In addition, this panel includes vascular abnormalities, such as dolichoectasia and hereditary hemorrhagic telangiectasia. Hereditary heart diseases are relatively rare, but they can lead to sudden cardiac death. For example, estimates suggest that people with short QT syndrome have a 40% chance of experiencing sudden cardiac death by age 40, with the highest amount of risk between ages 0 to 1 and 20 to 40 years old (Rudic et al., 2014).

The manifestation time depends on the particular disease but usually occurs in childhood or adulthood between 30 and 40 years (Gourraud et al., 2017; Dewi & Dharmadjati, 2020).

Coronary artery disease is a global health concern. It affects millions of individuals across diverse populations, ethnicities, and countries. The risk of developing CAD increases with age. It is more commonly observed in individuals over the age of 50 (Bauersachs et al., 2019). Heritability of death from coronary heart disease is estimated to be 0.57 (95% CI, 0.45-0.69) amongst male twins and 0.38 (0.26-0.50) amongst female twins, which was shown by cohort study with follow-up (Zdravkovic et al., 2002).

Thrombophilia is a group of genetic and acquired disorders that increase an individual's tendency to develop abnormal blood clots, known as thrombosis. These clots can occur in veins (venous thrombosis) or arteries (arterial thrombosis) and can have serious health implications. Venous thromboembolism (VTE) ranks as the third leading global cause of mortality. The frequency of VTE occurrence displays variability across nations, spanning from 1-2 cases per 1000 person-years in Western Countries, whereas, in Eastern Countries, the incidence stands lower (<1 per 1000 person-years). The most common risk factors for VTE include male sex, diabetes, obesity, smoking, and genetic predisposition (Pastori et al., 2023).

We have added to this report testing 8 most relevant SNPs, which are tightly associated with an increased risk of developing thrombotic complications, like Factor V Leiden, Prothrombin G20210A Gene Mutation, Plasminogen Activator Inhibitor-1 mutation (see Supplementary for the whole list of SNPs). This section also includes PRS estimation for a blood clot or deep vein thrombosis (PGS000931) with 534 gene variants included.

The cumulative incidence of VTE increases with age in men and women and grows exponentially after the age of 50 (Arnesen et al., 2021).

Imbalances in the processing, transportation, or utilisation of lipids (fats) within the body lead to lipid metabolism disorders. Abnormalities in lipid metabolism can have significant health implications and are often associated with conditions such as cardiovascular diseases, diabetes, and metabolic syndrome. Dyslipidemia is a predictor of cardio-metabolic disease, including obesity. Genome-wide association studies have identified over 150 loci related to abnormal lipid levels (Matey-Hernandez et al., 2018). Hyperlipidemias usually do not cause direct clinical symptoms, so they often remain undiagnosed until a serious cardiovascular event occurs. There are well-established treatment options available to prevent the occurrence of atherosclerosis, so early diagnostics and risk estimation are essential (Rosada et al., 2020). Chen et al. (2023) have recently shown that Genetically proxied LDLR variants, which mimic the effects of lowering low-density lipoprotein cholesterol (LDL-C), were associated with extended lifespan.

Based on findings from the National Health and Nutrition Examination Survey (NHANES), it was observed that elevated low-density lipoprotein cholesterol (LDL-C) levels were present in 11.7% of adults aged 20–39 and 41.2% of adults aged 40–64. However, treatment rates were relatively lower, with only 10.6% of adults aged 20–39 and 47.7% of adults aged 40–64 receiving treatment for hyperlipidemia (Navar-Boggan et al., 2015).

## Longevity-enhancing ("pro-longevity") gene variants and PRSs

This category includes variants believed to provide protection against various diseases and play a role in fundamental ageing mechanisms. Despite the strong hereditary component of longevity, the exact role of many genes in either increasing or decreasing the chances of reaching the age of 100+ remains unknown. However, in the past three decades, ageing research has gone beyond simply spotting traits associated with ageing. Although a significant portion of the genes associated with longevity in model organisms has not been replicated in human studies, the metabolic pathways involved in the ageing process are evolutionarily conserved. Ukraintseva et al. (Ukraintseva et al., 2021) suppose that interplay between genes in these pathways may influence human lifespan. This deeper understanding of how genes affect specific metabolic pathways is essential for developing precise therapies aimed at addressing the ageing process (Campisi et al., 2019, Mutlu et al., 2021)

So, in order to help users better understand the role of gene variants in longevity, we have categorised all gene variants in the report into 11 distinct groups according to the longevity pathways they contribute to. These groups include genes associated with lipid transfer and signalling, the insulin/IGF-1 signalling pathway, antioxidant defence, mitochondrial function, the sirtuin and mTOR pathways, tumour-suppressor genes, the renin-angiotensin system, heat-shock protein genes, inflammation, and related pathways, and genome maintenance and post-transcriptional processes.

It is important to note that even though the functions of many genes involved in these metabolic pathways are not thoroughly understood in terms of their impact on longevity, gene polymorphisms in these pathways often act as risk factors for age-related diseases. Therefore, information about them can be valuable for the development of a set of preventive measures.

**Lipid transfer and signalling pathway.** Lipids play crucial roles in regulating ageing and longevity. Lipids are key biological molecules contributing to cellular and organismal functions in three principal ways. First, they are fundamental structural elements of cellular membranes. Second, they are key molecules in energy metabolism to fuel the cell. Third, they play roles by acting as signalling molecules. Ageing is associated with changes in lipid metabolism (Mutlu et al., 2021).

Lipid metabolism is not considered a separate longevity pathway, but genes that regulate lipid transfer, like APOE (Sebastiani et al., 2018) and CETP (Soerensen et al., 2012), show the strongest association with longevity.
APOE ε4 is associated with substantially decreased odds for extreme longevity. These gene variants are also associated with various age-related diseases. APOE ε4 allele remains the strongest genetic risk factor for sporadic Alzheimer's disease (Serrano-Pozo et al., 2021). Even so, no therapies directed at APOE are currently available, phase 2 clinical trials of gene therapy are ongoing (CTG Labs, n.d.) and the strategies to target APOE based on antisense oligonucleotides, monoclonal antibodies, and gene/base editing are developing (Yang et al., 2021).

Decreased CETP activity is associated with cardiovascular health and maintenance of good cognitive performance (Oestereich et al., 2022). CETP inhibitors are undergoing clinical trials and are showing promising results (Cupido et al., 2022, Khomtchouk et al., 2023).

Other genes included in this section (APOC3, APOC1, PON1, PPARG) contribute to the risk of cardiovascular disease development and dyslipidemia.

So, information about gene variants in lipid transfer pathways can also be a basis for proactive steps in the development of disease prevention strategies.

**Insulin/Insulin-like Growth Factor 1 (IGF-1) signaling pathway**. Studies have clearly shown that this pathway is conserved and involved in the ageing process in humans, mice, *C. elegans*, and *Drosophila* (Khan et al., 2019). Insulin regulates blood sugar levels by allowing cells to take in glucose, while Insulin-like Growth Factor 1 (IGF-1) promotes cell growth and development, playing a vital role in overall growth and tissue repair in humans. Gathering evidence indicates that this pathway has a crucial part in the development of various age-related conditions, including cancer, dementia, cardiovascular issues, and metabolic diseases As people age, their glucose metabolism is disrupted, leading to a range of diseases. At the same time, long-lived individuals exhibit improved insulin sensitivity. This has led to active research into the GH/IGF-I/insulin system in long-lived individuals (Vitale et al., 2017). GWAS studies have shown a clear association between genetic variation of genes involved in insulin/IGF-1 pathway and human longevity (Deelen et al., 2011).
This pathway plays a crucial role in mediating the effects of calorie and dietary restrictions. Furthermore, a significant portion of drugs considered potential geroprotectors (such as metformin and calorie restriction mimetics) specifically target this biochemical pathway. Authors (Vitale et al., 2019) suppose that understanding of the link between IGF-1/insulin system and longevity may have future clinical applications in promoting healthy ageing.
Thus, understanding the gene polymorphisms that impact the IGF-I/insulin pathway holds potential interest in assessing the risk of age-related diseases and selecting strategies for geroprotectors. In the report, we have included an analysis of 10 genes, products of which are involved in that process: FOXO3, FOXO3A, IGF2R, IGF1R, INSR, IRS2, IGF2, GHRHR, INS, Klotho.

**Antioxidant defense and mitochondrial function**. These sections are quite closely related since mitochondrial dysfunction leads to increased reactive oxygen species (ROS) formation. Mitochondrial dysfunction is one of the hallmarks of ageing, leading to decreased energy production, sterile inflammation, and oxidative stress (Amorim et al., 2022). Both nuclear and mitochondrial genomes regulate mitochondrial function. Polymorphisms in mitochondrial and nuclear genes are studied in the context of longevity. For example, the m.8473T>C mutation is investigated for its contribution to the longevity of the Japanese population (Shamanskiy et al., 2019). The report includes polymorphisms of nuclear genes only. These single-point mutations affect the slowing down of the electron transport chain, reducing free radical production, the functioning of the first complex of the respiratory chain, mitochondrial fission processes, and mitochondrial biogenesis. The report includes an analysis of gene polymorphisms: MTPP, UCP3, UCP2, NDUFS1, PPARG, TOMM40, HIF1A.

Antioxidant defense plays an important role in the ageing process and longevity. Oxidative stress, which is caused by an imbalance between the production of reactive oxygen species (ROS) and the body's ability to neutralize them, is a major contributor to age-related diseases and the ageing process. The body has various mechanisms to defend against oxidative stress, including antioxidant enzymes such as superoxide dismutase (SOD), catalase, and glutathione peroxidase, as well as non-enzymatic antioxidants such as vitamins C and E, and glutathione. These antioxidants work together to neutralize ROS and prevent damage to cells and tissues.

Studies have shown that increased antioxidant defence can promote longevity and delay the onset of age-related diseases (Finkel & Holbrook, 2000; Dato et al., 2013). This is a list of genes we analyse related to antioxidant defense: SOD1, SOD2, GSTZ1-1, GSR, TXNRD1.

**Genes related to the sirtuin pathway.** The sirtuin genes (SIRT1-SIRT7) are a family of genes that are involved in regulating cellular processes such as DNA repair, metabolism, and stress response. The sirtuin pathway, which involves the activity of these genes, has been implicated in regulating longevity (Watroba & Szukiewicz, 2021). Studies have shown that activation of sirtuins can increase lifespan in several model organisms, including yeast, worms, and mice (Ledford, 2012; Wierman & Smith, 2013; Guarente, 2007). This is thought to be due to the role of sirtuins in promoting cellular resilience and protecting cells from damage.

In humans, variations in the SIRT genes have been associated with age-related diseases such as Alzheimer's disease, cardiovascular disease, and cancer (Zhao et al., 2020). Some studies have also suggested that increased activity of SIRT genes may be associated with increased lifespan and improved healthspan in humans. Pharmacological Sirt1 activators are under clinical development (Dai et al., 2018). Various products and natural products, so-called 'Sirtfoods" are also studied in terms of their ability to activate sirtuins (Akan et al., 2022). This is a list of genes included in the report, related to the sirtuin pathway: SIRT6, SIRT1, SIRT3.

**MTOR,** also known as the mechanistic target of rapamycin, is a gene that encodes for a protein kinase involved in multiple cellular processes, including growth, metabolism, and ageing. The mTOR pathway plays a complex role in ageing and longevity. On the one hand, activation of the mTOR pathway has been shown to promote cellular growth and proliferation, which may be beneficial for tissue repair and regeneration during early life stages. On the other hand, chronic activation of the mTOR pathway has been implicated in age-related diseases such as cancer, metabolic disorders, and neurodegeneration (Papadopoli et al., 2019).

Studies have shown that genetic or pharmacological inhibition of the mTOR pathway can extend the lifespan in a variety of model organisms, including mice, flies, and worms. In recent times, there has been a growing focus among basic and translational scientists, clinicians, and biotechnology companies on the targeted inhibition of mTORC1 as a potential treatment for age-related conditions. Based on studies in rodents and humans, mTOR inhibitors may have potential benefits for ameliorating or slowing age-related conditions associated with the brain, the heart, the liver, skeletal muscle, tendons, the immune system, the skin and the intestine (Mannick & Lamming, 2023).

Genetic variants in the MTOR gene may be associated with differences in lifespan (Passtoors et al., 2012) and age-related disease risk (Yates et al., 2013), cancer (Min et al., 2022) and, potentially, can influence the efficiency of rapamycin and rapalogs.

**Tumour-suppressor genes**. Tumour suppressor genes and cell cycle regulators are not typically considered as "longevity genes" per se, but they do play an important role in the ageing process and the development of age-related diseases, including cancer.

One well-known tumour suppressor gene is TP53, which has been shown to play a role in regulating cellular senescence and preventing the accumulation of damaged cells. Its role in longevity is twofold. On one hand, by preventing tumours early in life, p53 is an essential gene for ensuring longevity. On the other hand, sustained p53 activation has the potential to reduce the self-renewal capacity of stem/progenitor cells, which could lead to age-related consequences (Feng et al., 2011). Similarly, cell cycle regulators such as cyclin-dependent kinases (CDKs) are involved in regulating the timing and progression of cell division. The dysregulation of CDKs has been implicated in a variety of age-related diseases, including cancer and neurodegenerative disorders (Łukasik et al., 2021). Into the report, we have included the estimation of polymorphisms in the following genes: TP53
ATM, CDKN2B, CDKN2B-AS1, CDKN2B-AS1, CDKN1A, CDK14, CDK6.

**Genes of the renin-angiotensin system**. The renin-angiotensin system (RAS) plays an important role in the regulation of blood pressure, fluid balance, and electrolyte homeostasis. However, it is also involved in the ageing process and the development of age-related diseases. One of the mechanisms by which RAS influences ageing is through the activation of oxidative stress and inflammation. RAS also affects the cardiovascular system and is involved in the development of hypertension, a major risk factor for cardiovascular diseases such as heart attack and stroke (Zhu et al., 2003). Furthermore, RAS activation has been linked to the development of insulin resistance and metabolic syndrome, which are also associated with ageing and age-related diseases (Rahimi, 2016; Wang et al., 2010). In addition, RAS has been implicated in the regulation of cellular senescence, a process by which cells lose their ability to divide and contribute to ageing and age-related diseases (Mogi, 2020). We have included in the report following genes, contributing to RAS work: ACE ,AGTR1, AGT.

**Heat-shock protein genes**. The HSP (heat shock protein) genes are a group of genes that encode heat shock proteins, which are a class of chaperone proteins that help to protect cells from stress-induced damage. HSPs have been shown to play a role in a variety of cellular processes, including protein folding, DNA repair, and apoptosis. While the exact mechanisms by which HSPA genes influence ageing and longevity are not yet fully understood, it is believed that they may help to protect cells from the accumulation of damage caused by stress and other environmental factors. As such, HSPs and their associated genes may be a promising target for interventions aimed at promoting healthy ageing and extending lifespan (Peinado-Ruiz et al., 2022). Some unfavourable variants in HSP genes have been shown to reduce the ability of cells to respond to stress, leading to increased damage and inflammation.Report contain the data regarding polymorphisms found in HSPA1L, HSPA1A, HSPA1B, HSPA14 genes.

**Inflammation and related pathways.** Chronic inflammation is a major contributor to the ageing process. Chronic inflammation is characterised by the sustained activation of the immune system and the release of pro-inflammatory molecules, such as cytokines, chemokines, and reactive oxygen species (ROS). These molecules can damage cellular components, including DNA, proteins, and lipids, leading to cellular dysfunction and death (X. Li et al., 2023).

In addition, chronic inflammation can also activate other pathways involved in ageing, such as the mTOR pathway and the senescence-associated secretory phenotype (SASP), which further exacerbate inflammation and tissue damage. Therefore, reducing chronic inflammation is a promising strategy to promote healthy ageing and prevent age-related diseases.Chronic inflammation can have a genetic component, with certain genes being associated with increased inflammation levels (Newcombe et al., 2018; Ellinghaus et al., 2016). Report includes the information regarding genetic variants in CCL5, IL6, IL10, TNF, VEGFA, TLR4.

Multiple strategies are developing to tackle chronic inflammation, including diet interventions (Galland, 2010), pharmacological NLRP3 inflammasome inhibitors (Zahid et al., 2019), senolytic drugs (Hernández-Silva et al., 2022) etc.

**Genome maintenance and post-transcriptional processes.** Genome maintenance and post-transcriptional processes are both important for maintaining cellular and organismal homeostasis, and both have been implicated in the regulation of longevity. Genome maintenance is essential for preventing DNA damage and mutations that can lead to age-related diseases and shortened lifespans (Cho & Suh, 2014). Post-transcriptional processes, including RNA splicing, translation, and decay, are important for regulating gene expression and ensuring that proteins are produced at the appropriate levels and times. Dysregulation of these processes can lead to age-related diseases and shortened lifespans (Santos & Lindner, 2017).

The report includes information regarding two genes involved in that process, WRN and ADARB1, since they were shown to affect lifespan.

## Pharmacogenetics of longevity drugs

Genetic variations can cause increased sensitivity or resistance to drugs, leading to more or fewer adverse side effects and potentially altering the overall efficacy of the treatment. Many pharmaceutical drugs, such as statins, sartans, and anticoagulants, are taken by elderly individuals. These medications significantly aid in reducing the risk of mortality from cardiovascular diseases and diabetes complications (Horodinschi et al., 2019; Wu et al., 2022; S.-H. Kim et al., 2013). However, the effectiveness of these drugs and the risk of side effects are largely influenced by genetic characteristics (Guan et al., 2019; S.-H. Kim et al., 2013). Adverse effects can affect both the individual's health and lifespan. Moreover, some drugs, like metformin and rapamycin, are potentially pro-longevity. However, there also can be personalised risks and a need for dosage correction according to genotype.

Our report currently provides information on the following drugs: acamprosate, allopurinol, bupropion, carbamazepine, Imatinib, liraglutide, metformin, pioglitazone, rivastigmine, sunitinib, and warfarin. Among these, metformin, pioglitazone, and liraglutide are noteworthy due to their strong association with longevity pathways. Additionally, we are interested in drugs prescribed to older individuals, as they can potentially lead to adverse effects, particularly when considering their pharmacogenetics.

Metformin is a first-line agent for the treatment of type 2 diabetes. Metformin displays a promising perspective in ageing-related clinical applications. The addition of metformin to the diet can delay ageing and increase life span in both *C. elegans* and rodents (Cabreiro et al., 2013; De Haes et al., 2014). The longevity pathways affected by metformin include: activation of AMP-activated kinase (AMPK), mTOR inhibition, mitohormesis, and respiratory chain Complex I inhibition.

A multicenter, randomised, double-blind, placebo-controlled clinical trial, Targeting Aging with Metformin (TAME), has been initiated to further explore the anti-aging role of metformin (Justice et al., 2018). A systematic review and meta-analysis data showed that the use of metformin results in a reduction of all-cause mortality associated with diseases that accelerate ageing, including cancer and cardiovascular disease. Metformin use leads to weight reduction and reduces HbA1c; metformin also lowers the atherosclerotic burden in non-diabetic persons at risk of developing T2DM (Mohammed et al., 2021). However, the stratification of responders and non-responders is essential and can also minimise the risks of adverse effects (Chen et al., 2022). It may cause lactic acidosis mitochondrial dysfunction; several studies have reported that the chronic use of metformin results in vitamin B12 deficiency, affecting between 6 and 30% of users (J. Kim et al., 2019). Early GWAS studies identified that the combined impact of the ATM and SLC2A2 may influence the metformin response; even so, the effect was minimal (Florez, 2017). More recent studies showed that rs12208357 in the SLC22A1 gene had a significant impact on response to metformin in T2DM patients (Nasykhova et al., 2022).

Liraglutide is a glucagon-like peptide-1 (GLP-1) receptor agonist (RA). GLP-1RAs comprise a class of peptides that bind and activate receptors for the endogenous incretin hormone GLP-1. They potentiate insulin secretion and reduce glucagon secretion in a glucose-dependent manner to lower blood glucose. Anti-diabetic medication treats type 2 diabetes, obesity, and chronic weight management. It was shown to have multiple positive health effects for the elderly. In a prospective case series study liraglutide has shown efficacy in the treatment of obesity in the elderly, fat loss, and safety to prevent sarcopenia (Perna et al., 2016). Lira added to insulin therapy may improve the quality of life in elderly T2D patients undergoing insulin therapy (Tonoike et al., 2018). Multiple rodent studies show that Liraglutide ameliorates cognitive dysfunction, improves the cognitive function of diabetic mice (H. Zhang et al., 2020), and alleviates cognitive deficit (M. Zhang et al., 2021). It may positively affect Alzheimer's disease treatment (Vargas-Soria et al., 2021). Potentially, it can be used to extend lifespan since it downregulates receptors of advanced glycation end products (RAGE) associated with fibrosis, inflammation, and cognition decline. However, it also can cause adverse effects, like an increase of the heart rate of T2DM patients (Seo, 2021).

Pioglitazone is also an anti-diabetic drug and insulin-sensitizer from the thiazolidinedione (TZD) class. It activates peroxisome proliferator-activated receptor-γ (PPARγ). PPARγ agonists are promising in delaying ageing and enhancing longevity since PPARγ activation suppresses age-induced inflammation and oxidative stress (Xu et al., 2020). Retrospective analysis of the effects of the PPARγ agonist pioglitazone (Pio) on longevity showed decreased mortality in patients receiving Pio compared to those receiving a PPARγ-independent insulin secretagogue glimepiride. Taken together, these data suggest the possibility of using PPARγ agonists to promote healthy ageing and extend lifespan (Xu et al., 2020). Pioglitazone is an effective therapy for elderly patients with type 2 diabetes mellitus (Rajagopalan et al., 2004). In rodent models it has been shown, that pioglitazone attenuates ageing-related disorders (reduced atherosclerosis, hepatic steatosis, renal ageing, and skin atrophy) in aged apolipoprotein E deficient mice (Shen et al., 2018). However, pioglitazone-induced heart failure is known in patients with underlying heart disease, with rare cases documented for patients with normal left ventricular function (Jearath et al., 2016). Up to now, the genes CYP2C8 and PPARG have received the most comprehensive research, and specific genetic variations have been identified as factors contributing to the variability in both the way pioglitazone is processed in the body (pharmacokinetics) and its effects (pharmacodynamics). Notably, individuals with the PPARG 12Ala variant experienced a more positive change in fasting blood glucose levels compared to those with the common Pro12Pro genotype, as supported by a p-value of 0.018 (Kawaguchi-Suzuki & Frye, 2013).

Rivastigmine (sold under the trade name Exelon, among others) is a dual inhibitor of both acetylcholine esterase (AChE) and butyrylcholinesterase (BuChE), enzymes involved in the hydrolysis of acetylcholine, used for the treatment of mild to moderate Alzheimer's disease. The drug can be administered orally or via a transdermal patch (Khoury et al., 2018). Several large, placebo-controlled, double-blind trials have demonstrated that using rivastigmine results in significant improvements in the cognitive, functional, and global performances of AD patients (Annicchiarico et al., 2007). It was recently shown that rivastigmine modifies the α-secretase pathway and potentially early Alzheimer's disease, so it is a disease-modifying intervention that goes beyond symptomatic treatment for AD (B. Ray et al., 2020). However, long-term rivastigmine treatment increases all-cause mortality in patients with Alzheimer's disease (Kazmierski et al., 2018). IL-6 and A2M gene polymorphisms were proposed to impact the efficacy of rivastigmine on AD patients (Zamani et al., 2015).

Warfarin is a widely prescribed anticoagulant. Warfarin has a narrow therapeutic window and a large variation in dose requirements from one patient to another, and this is partly due to genetic variations. Key genes involved in warfarin pharmacogenetics include CYP2C9 and VKORC1. The FDA (U.S. Food and Drug Administration) recognizes the significance of warfarin pharmacogenetics and has included genetic information in the drug's label to guide dosing (Drozda et al., 2018). Besides direct adverse effects, warfarin can cause delayed ones, affecting longevity, since it may cause vascular calcification, which is a risk factor for CVDs (Palaniswamy et al., 2011).

So, testing pharmacogenetics is essential to understanding individual metabolism, identifying personalised risks, optimising drug treatments, and promoting better health outcomes. We plan to expand our pharmacogenetics report by including more medications that have already received FDA approval or are in clinical trials to treat major age-related diseases or age-related changes.

## Unresolved problems

### Population estimation

The genetic variants that are associated with longevity may differ between populations due to differences in genetic backgrounds, environmental exposures, and lifestyle factors. Many studies on longevity genetics have been conducted on populations of European descent, which limits the generalizability of findings to other ethnic groups. There are just a few gene variants that are validated across multiple populations (Deelen et al., 2019). Moreover, populations are defined in different ways in various studies. As a result, using the same set of longevity variants across all populations may lead to inaccurate or biassed results.

To address this issue, developers of personalised genomic platforms must consider population diversity when designing and interpreting their analyses. It can be done by implementing human ancestry identification tools. The population to which the user belongs can be computed in different ways. Nevertheless, distinguishing between individuals from closely associated sub-populations (e.g., from the same continent) is still a difficult challenge (Toma et al., 2018). Some research suggests the possibility to estimate biogeographical ancestry by analysing SNP panels derived from just one chromosome (Toma et al., 2018), while others use multiple SNP markers throughout the genome. Numerous software tools were developed, such as STRUCTURE, ADMIXTURE, FastPop, GrafPop, etc., for estimating ancestry in a model-based large autosomal SNP genotype datasets (Lawson et al., 2018). They utilise different quantities of SNP markers, from 100 up to 10,000 SNPs (Sampson et al., 2011; Jin et al., 2019; Li et al., 2016). The general trend is minimising the subset of SNPs that can predict ancestry with a minimal error rate (Sampson et al., 2011).

An array of private companies, such as 23andMe and AncestryDNA, provide direct-to-consumer (DTC) genetic testing by analysing ancestry informative markers to determine geographic origins.

So far, we do not automatically pre-filter the variant results based on users' population letting the users design which studies they consider important for themselves. We are currently working on choosing the most appropriate software tool to implement an efficient human ancestry identification tool for Just-DNA-Seq platform.

## Conclusion

In a world where personal genomics is revolutionising medicine, longevity genetics stands out as a promising frontier. It empowers individuals with the knowledge of how their unique genetic makeup influences health and lifespan. With increasing accessibility to genomic data and evolving research, personal genomics offers a pathway to better disease prevention, management, and overall well-being.

Understanding the heritability of longevity remains an ongoing exploration, with multiple genetic variants at play. FOXO3 and APOE are among the candidates associated with longevity, and rare coding variants continue to add depth to this field.

To facilitate the translation of research findings into practical scientific applications, platforms like Just-DNA-Seq have become indispensable. Just-DNA-Seq, an open-source genomic platform uniquely tailored to longevity research, enables researchers and users to upload VCF files and receive annotations of genetic variants and polygenic risk scores. These insights are organized into reports, each holding significance for health and longevity across different stages of ontogenesis. Just-DNA-Seq offers reports on rare inherited diseases, hereditary cardiovascular conditions, risks associated with age-related diseases, genetic variants linked to exceptional longevity, and individual drug metabolism profiles. These features are particularly relevant for scientific inquiries concerning older individuals. Our open-source personal genomics solutions prioritize transparency and privacy, empowering researchers to utilize genetic data for scientific advancement in health and longevity research.

## Contribution statements:

Kulaga Anton - co-founded the project, participated in bioinformatic pipeline development, overall architecture development, team management, article writing
Borysova Olga - modules and report content, database curation, setting appropriate filters, developing the report structure, data collection and analysis, interpreting the results. Additionally, was actively involved in drafting and revising the manuscript, ensuring clarity and accuracy in presenting the research findings.
Karmazin Alexey - lead architect of the report project, participated in writing OakVar report modules and preparation utiles, writing PRS modules.
Maria Koval - participating in developing report modules, report template, and writing health risks modules and code availability parts of the article
Nikolay Usanov - co-founded the project, participated in writing bioinformatic pipelines, fundraising
Fedorova Alina - bioinformatic utilities, ML module, scientific literature exploration
Pushkareva Malvina - software development and testing
Evfratov Sergey - developed drug module
Ryangguk Kim - created and maintained OakVar, extended OakVar with additional features required for Just-DNA-Seq project
Tacutu Robi - bioinformatics and ageing research advising

Gratitudes:
* George Fuellen for valuable suggestions and comments
* Volodymir Semenuik for help with documentation

## Code availability

Links to our sources:

There is a GitHub organisation where all the source code and additional programs that were created during the work are stored: https://github.com/dna-seq

The organisation contains the following types of repositories:

## Competing Interests:

Ryangguk Kim is a co-founder of OakVar Inc. which provides additional services on top of open-source OakVar platform
Anton Kulaga, Olga Borysova, Maria Koval, Nikolay Usanov and Alex Karmazin are co-founders of SecvADN SRL. company that provides additional services on top of open-source modules

OakVar Just-DNA-Seq modules:

https://github.com/dna-seq/oakvar-longevity - a repository that contains all the modules and the longevity reporter.
https://github.com/dna-seq/longevity2reporter - Longevity2reporter displays your genome analysis related to longevity and disease risks.
https://github.com/dna-seq/just_longevitymap - Longevity map postaggregator for longevity report in OakVar.
https://github.com/dna-seq/just_prs - Polygenic risk score (PRS) postaggregator for longevity report in OakVar.
https://github.com/dna-seq/just_drugs - Drugs postaggregator for longevity reporter in OakVar.
https://github.com/dna-seq/just_cancer - Cancer postaggregator for longevity reporter of Oakvar.
https://github.com/dna-seq/just_cardio - Cardio postaggregator for longevity report.
https://github.com/dna-seq/just_coronary - Coronary disease postaggregator for longevity report in OakVar.
https://github.com/dna-seq/just_lipidmetabolism - Lipid metabolism postaggregator for longevity report in OakVar.
https://github.com/dna-seq/just_thrombophilia - Thrombophilia risks postaggregator for longevity report in OakVar.

Postagregator repository consists of the following elements:

- data folder that contains data (sqlite3 database or a text file)
- source code (python program file, can be a few files)
- a logotype for OakVar GUI, yml file that contains settings for the proper integration of a module into OakVar
- .md file that contains a small description of a module for OakVar GUI
- README.md file that contains a small description of a module for GitHub
- licence file
- .gitignore file

Reporter repository consists of the following elements:

- source code (python program file)
- template.html file, which is used as a template to generate a report by a python program
- a logotype for OakVar GUI, yml file that contains settings for the proper integration of a module into OakVar
- .md file that contains a small description of a module for OakVar GUI
- README.md file that contains a small description of a module for GitHub
- licence file
- .gitignore file

Experimental modules that require further validation:

OakVar Superhuman genes module:

https://github.com/dna-seq/superhumanreporter - Reporter of superhuman genes module of Oakvar.
The report is created from the data that was generated by just_superhuman postaggregator.

https://github.com/dna-seq/just_superhuman - Superhuman genes postaggregator of superhumangenes report in Oakvar (based on It depends on annotators: dbsnp, clinvar, omim, ncbigene, pubmed, gnomad.

Bioinformatic pipelines:

https://github.com/dna-seq/dna-seq - from this repository, the project started. It includes WDL bioinformatic pipelines allowing to quality control, align, variant call and annotating users fastq files. Use of outdated reference genomes, namely unpatched *GRCh37* versions, raw data without QC and False Positive prone Variant Calling tools etc, results in poor quality of VCF files. Be it a case of a commercial company using outdated pipelines and tools or just the sequencing that had been done long ago, one of the important practical uses of this module is to "refresh" the Variant Calls (VCF) file using cutting-edge open-source tools and latest reference sequence data to improve the accuracy and fidelity of the analysed SNPs.

The repository contains dockerized Cromwell (https://github.com/broadinstitute/cromwell) server and client (https://github.com/antonkulaga/cromwell-client ) to run WDL pipelines.

To refresh the VCF file using raw reads data, one needs to go through the following steps:

1) Sometimes DNA-Sequencing companies don't provide raw FASTA read files, but provide aligned BAM files. In such case, disassemble one's genome BAM files back into RAW FASTQ using "bam_to_fastq.wdl" pipeline
2) Clean (quality control) and re-align the raw reads to the up-to-date Human Genome versions of choice in one go, using "alignment.wdl". To speed up the process, dockerized bwa-mem2 (https://github.com/bwa-mem2/bwa-mem2 ) tool is used for alignment, its output results are identical to renown BWA aligner tool.
3) Finally, get a new up-to-date VCF for further analysis by using "deep_variant.wdl" pipeline, utilise the power of Google's latest DeepVariant which continues to show the best accuracy and outperform the competition ("Pipelines based on DeepVariant consistently perform better than all other considered solutions"[#]).

#Barbitoff, Y.A., Abasov, R., Tvorogova, V.E. *et al.* Systematic benchmark of state-of-the-art variant calling pipelines identifies major factors affecting accuracy of coding sequence variant discovery. *BMC Genomics* **23**, 155 (2022). https://doi.org/10.1186)

https://github.com/dna-seq/neural_networks a framework for Interpretable Neural Networks for Genetics.

Websites and documentation:

https://dna-seq.github.io/ is an official static website of the project backed by https://github.com/dna-seq/dna-seq.github.io

https://dna-seq.github.io/report_example/ is an example of a report that is generated my Just-DNA-Seq longevity module backed by https://github.com/dna-seq/report_example

https://just-dna-seq.readthedocs.io/en/oakvar/index.html  Just-DNA-Seq documentation backed by https://github.com/dna-seq/just-dna-seq

Auxiliary repositories:

Other repositories in the github organisation, mostly used for data wrangling, docker containers, and other purposes

https://github.com/dna-seq/my_genes_and_longevity_comic a repository that contains a comic based on Just-DNA-Seq project

https://github.com/dna-seq/containers a repository for containers in Just-DNA-Seq project.

https://github.com/dna-seq/modules-preprocessing a repository with the code for preprocessing data for Just-DNA-Seq modules

https://github.com/dna-seq/longevity-map source code for a website which can be used to find the information about different human genetic variants associated with longevity (based on LongevityMap by João Pedro group)

https://github.com/dna-seq/stroke_prs and https://github.com/dna-seq/prs - old experimental repositories for PRS module

https://github.com/dna-seq/gero-drugs-module - old experimental repository for Drug module

https://github.com/dna-seq/longevity-annotator - an archive of an old version of Just-DNA-Seq modules

# Bibliography

Akan, O. D., Qin, D., Guo, T., Lin, Q., & Luo, F. (2022). Sirtfoods: New concept foods, functions, and mechanisms. *Foods*, *11*(19), 2955. https://doi.org/10.3390/foods11192955

Amorim, J. A., Coppotelli, G., Rolo, A. P., Palmeira, C. M., Ross, J. M., & Sinclair, D. A. (2022). Mitochondrial and metabolic dysfunction in ageing and age-related diseases. *Nature Reviews Endocrinology*, *18*(4), 243–258. https://doi.org/10.1038/s41574-021-00626-7

Annicchiarico, R., Federici, A., Pettenati , C., & Caltagirone, C. (2007). Rivastigmine in Alzheimer's disease: Cognitive function and quality of life. *Therapeutics and Clinical Risk Management*, *3*(6), 1113–1123.

Arnesen, C. A. L., Veres, K., Horváth-Puhó, E., Hansen, J.-B., Sørensen, H. T., & Brækkan, S. K. (2021). Estimated lifetime risk of venous thromboembolism in men and women in a Danish nationwide cohort: Impact of competing risk of death. *European Journal of Epidemiology*, *37*(2), 195–203. https://doi.org/10.1007/s10654-021-00813-w

Auton, A., Abecasis, G. R., Altshuler, D. M., (Co-Chair), Durbin, R. M., (Co-Chair), Abecasis, G. R., Bentley, D. R., Chakravarti, A., Clark, A. G., Donnelly, P., Eichler, E. E., Flicek, P., Gabriel, S. B., Gibbs, R. A., Green, E. D., Hurles, M. E., Knoppers, B. M., Korbel, J. O., Lander, E. S., Lee, C., … Abecasis, G. R. (2015). A global reference for human genetic variation. *Nature*, *526*(7571), 68–74. https://doi.org/10.1038/nature15393

Bauersachs, R., Zeymer, U., Brière, J.-B., Marre, C., Bowrin, K., & Huelsebeck, M. (2019). Burden of coronary artery disease and peripheral artery disease: A literature review. *Cardiovascular Therapeutics*, *2019*, 1–9. https://doi.org/10.1155/2019/8295054

Bin-Jumah, M. N., Nadeem, M. S., Gilani, S. J., Al-Abbasi, F. A., Ullah, I., Alzarea, S. I., Ghoneim, M. M., Alshehri, S., Uddin, A., Murtaza, B. N., & Kazmi, I. (2022). Genes and longevity of lifespan. *International*

Journal of Molecular Sciences*, 23*(3), 1499. https://doi.org/10.3390/ijms23031499

Bonomi, L., Huang, Y., & Ohno-Machado, L. (2020). Privacy challenges and research opportunities for genomic data sharing. *Nature Genetics*, *52*(7), 646–654. https://doi.org/10.1038/s41588-020-0651-0

Broer, L., Buchman, A. S., Deelen, J., Evans, D. S., Faul, J. D., Lunetta, K. L., Sebastiani, P., Smith, J. A., Smith, A. V., Tanaka, T., Yu, L., Arnold, A. M., Aspelund, T., Benjamin, E. J., De Jager, P. L., Eirkisdottir, G., Evans, D. A., Garcia, M. E., Hofman, A., … Murabito, J. M. (2014). GWAS of longevity in CHARGE consortium confirms APOE and FOXO3 candidacy. *The Journals of Gerontology: Series A*, *70*(1), 110–118. https://doi.org/10.1093/gerona/glu166

Brooks-Wilson, A. R. (2013). Genetics of healthy aging and longevity. *Human Genetics*, *132*(12), 1323–1338. https://doi.org/10.1007/s00439-013-1342-z

Browning, B. L., Zhou, Y., & Browning, S. R. (2018). *A one penny imputed genome from next generation reference panels*. Cold Spring Harbor Laboratory. http://dx.doi.org/10.1101/357806

Cabreiro, F., Au, C., Leung, K.-Y., Vergara-Irigaray, N., Cochemé, H. M., Noori, T., Weinkove, D., Schuster, E., Greene, N. D. E., & Gems, D. (2013). Metformin retards aging in c. elegans By altering microbial folate and methionine metabolism. *Cell*, *153*(1), 228–239. https://doi.org/10.1016/j.cell.2013.02.035

Campisi, J., Kapahi, P., Lithgow, G. J., Melov, S., Newman, J. C., & Verdin, E. (2019). From discoveries in ageing research to therapeutics for healthy ageing. *Nature*, *571*(7764), 183–192. https://doi.org/10.1038/s41586-019-1365-2

Cariaso, M., & Lennon, G. (2011). SNPedia: A wiki supporting personal genome annotation, interpretation and analysis. *Nucleic Acids Research*, *40*(D1), D1308–D1312. https://doi.org/10.1093/nar/gkr798

Caruso, C., Ligotti, M. E., Accardi, G., Aiello, A., Duro, G., Galimberti, D., & Candore, G. (2022). How important are genes to achieve longevity? *International Journal of Molecular Sciences*, *23*(10), 5635. https://doi.org/10.3390/ijms23105635

Chen, S., Gan, D., Lin, S., Zhong, Y., Chen, M., Zou, X., Shao, Z., & Xiao, G. (2022). Metformin in aging and aging-related diseases: Clinical applications and relevant mechanisms. *Theranostics*, *12*(6), 2722–2740. https://doi.org/10.7150/thno.71360

Cho, M., & Suh, Y. (2014). Genome maintenance and human longevity. *Current Opinion in Genetics & Development*, *26*, 105–115. https://doi.org/10.1016/j.gde.2014.07.002

Collister, J. A., Liu, X., & Clifton, L. (2022). Calculating polygenic risk scores (PRS) in UK biobank: A practical guide for epidemiologists. *Frontiers in Genetics*, *13*. https://doi.org/10.3389/fgene.2022.818574

*CTG Labs*. (n.d.). NCBI. Retrieved October 29, 2023, from https://clinicaltrials.gov/study/NCT03634007

Cupido, A. J., Reeskamp, L. F., Hingorani, A. D., Finan, C., Asselbergs, F. W., Hovingh, G. K., & Schmidt, A. F. (2022). Joint genetic inhibition of PCSK9 and CETP and the association with coronary artery disease. *JAMA Cardiology*, *7*(9), 955. https://doi.org/10.1001/jamacardio.2022.2333

Dai, H., Sinclair, D. A., Ellis, J. L., & Steegborn, C. (2018). Sirtuin activators and inhibitors: Promises, achievements, and challenges. *Pharmacology & Therapeutics*, *188*, 140–154. https://doi.org/10.1016/j.pharmthera.2018.03.004

Dato, S., Crocco, P., D'Aquila, P., de Rango, F., Bellizzi, D., Rose, G., & Passarino, G. (2013). Exploring the role of genetic variability and lifestyle in oxidative stress response for healthy aging and longevity. *International Journal of Molecular Sciences*, *14*(8), 16443–16472. https://doi.org/10.3390/ijms140816443

De Haes, W., Frooninckx, L., Van Assche, R., Smolders, A., Depuydt, G., Billen, J., Braeckman, B. P., Schoofs, L., & Temmerman, L. (2014). Metformin promotes lifespan through mitohormesis via the peroxiredoxin PRDX-2. *Proceedings of the National Academy of Sciences*, *111*(24). https://doi.org/10.1073/pnas.1321776111

Deelen, J., Evans, D. S., Arking, D. E., Tesi, N., Nygaard, M., Liu, X., Wojczynski, M. K., Biggs, M. L., van der Spek, A., Atzmon, G., Ware, E. B., Sarnowski, C., Smith, A. V., Seppälä, I., Cordell, H. J., Dose, J., Amin, N., Arnold, A. M., Ayers, K. L., … Murabito, J. M. (2019). A meta-analysis of genome-wide association studies identifies multiple longevity genes. *Nature Communications*, *10*(1). https://doi.org/10.1038/s41467-019-11558-2

Deelen, J., Uh, H.-W., Monajemi, R., van Heemst, D., Thijssen, P. E., Böhringer, S., van den Akker, E. B., de Craen, A. J. M., Rivadeneira, F., Uitterlinden, A. G., Westendorp, R. G. J., Goeman, J. J., Slagboom, P. E., Houwing-Duistermaat, J. J., & Beekman, M. (2011). Gene set analysis of GWAS data for human longevity highlights the relevance of the insulin/IGF-1 signaling and telomere maintenance pathways. *AGE*, *35*(1), 235–249. https://doi.org/10.1007/s11357-011-9340-3

Dewi, I. P., & Dharmadjati, B. B. (2020). Short QT syndrome: The current evidences of diagnosis and management. *Journal of Arrhythmia*, *36*(6), 962–966. https://doi.org/10.1002/joa3.12439

Donlon, T. A., Morris, B. J., Chen, R., Masaki, K. H., Allsopp, R. C., Willcox, D. C., Tiirikainen, M., & Willcox, B. J. (2017). Analysis of polymorphisms in 59 potential candidate genes for association with human longevity. *The Journals of Gerontology: Series A*, *73*(11), 1459–1464.

https://doi.org/10.1093/gerona/glx247

Drozda, K., Pacanowski, M. A., Grimstein, C., & Zineh, I. (2018). Pharmacogenetic labeling of fda-approved drugs. *JACC: Basic to Translational Science*, *3*(4), 545–549. https://doi.org/10.1016/j.jacbts.2018.06.001

Ellinghaus, D., Jostins, L., Spain, S. L., Cortes, A., Bethune, J., Han, B., Park, Y. R., Raychaudhuri, S., Pouget, J. G., Hübenthal, M., Folseraas, T., Wang, Y., Esko, T., Metspalu, A., Westra, H.-J., Franke, L., Pers, T. H., Weersma, R. K., Collij, V., … Franke, A. (2016). Analysis of five chronic inflammatory diseases identifies 27 new associations and highlights disease-specific patterns at shared loci. *Nature Genetics*, *48*(5), 510–518. https://doi.org/10.1038/ng.3528

Feng, Z., Lin, M., & Wu, R. (2011). The Regulation of Aging and Longevity: A New and Complex Role of p53. *Genes & Cancer*, *2*(4), 443–452. https://doi.org/10.1177/1947601911410223

Finkel, T., & Holbrook, N. J. (2000). Oxidants, oxidative stress and the biology of ageing. *Nature*, *408*(6809), 239–247. https://doi.org/10.1038/35041687

Florez, J. C. (2017). The pharmacogenetics of metformin. *Diabetologia*, *60*(9), 1648–1655. https://doi.org/10.1007/s00125-017-4335-y

Galland, L. (2010). Diet and inflammation. *Nutrition in Clinical Practice*, *25*(6), 634–640. https://doi.org/10.1177/0884533610385703

Gourraud, J.-B., Barc, J., Thollet, A., Le Marec, H., & Probst, V. (2017). Brugada syndrome: Diagnosis, risk stratification and management. *Archives of Cardiovascular Diseases*, *110*(3), 188–195. https://doi.org/10.1016/j.acvd.2016.09.009

Guan, Z., Wu, K., Li, R., Yin, Y., Li, X., Zhang, S., & Li, Y. (2019). Pharmacogenetics of statins treatment: Efficacy and safety. *Journal of Clinical Pharmacy and Therapeutics*, *44*(6), 858–867. https://doi.org/10.1111/jcpt.13025

Guarente, L. (2007). Sirtuins in aging and disease. *Cold Spring Harbor Symposia on Quantitative Biology*, *72*(1), 483–488. https://doi.org/10.1101/sqb.2007.72.024

Hernández-Silva, D., Cantón-Sandoval, J., Martínez-Navarro, F. J., Pérez-Sánchez, H., de Oliveira, S., Mulero, V., Alcaraz-Pérez, F., & Cayuela, M. L. (2022). Senescence-Independent anti-inflammatory activity of the senolytic drugs Dasatinib, Navitoclax, and Venetoclax in zebrafish models of chronic inflammation. *International Journal of Molecular Sciences*, *23*(18), 10468. https://doi.org/10.3390/ijms231810468

Horodinschi, R.-N., Stanescu, A. M. A., Bratu, O. G., Pantea Stoian, A., Radavoi, D. G., & Diaconu, C. C.

(2019). Treatment with statins in elderly patients. *Medicina*, *55*(11), 721. https://doi.org/10.3390/medicina55110721

Hu, D., Li, Y., Zhang, D., Ding, J., Song, Z., Min, J., Zeng, Y., & Nie, C. (2022). Genetic trade‐offs between complex diseases and longevity. *Aging Cell*, *21*(7). https://doi.org/10.1111/acel.13654

Jearath, V., Vashisht, R., Rustagi, V., Raina, S., & Sharma, R. (2016). Pioglitazone-induced congestive heart failure and pulmonary edema in a patient with preserved ejection fraction. *Journal of Pharmacology and Pharmacotherapeutics*, *7*(1), 41–43. https://doi.org/10.4103/0976-500x.179363

Jin, Y., Schaffer, A. A., Feolo, M., Holmes, J. B., & Kattman, B. L. (2019). GRAF-pop: A fast distance-based method to infer subject ancestry from multiple genotype datasets without principal components analysis. *G3 Genes|Genomes|Genetics*, *9*(8), 2447–2461. https://doi.org/10.1534/g3.118.200925

Juckett, D. A. (2009). What determines age-related disease: Do we know all the right questions? *AGE*, *32*(2), 155–160. https://doi.org/10.1007/s11357-009-9120-5

Justice, J. N., Niedernhofer, L., Robbins, P. D., Aroda, V. R., Espeland, M. A., Kritchevsky, S. B., Kuchel, G. A., & Barzilai, N. (2018). Development of clinical trials to extend healthy lifespan. *Cardiovascular Endocrinology & Metabolism*, *7*(4), 80–83. https://doi.org/10.1097/xce.0000000000000159

Kawaguchi-Suzuki, M., & Frye, R. F. (2013). Current clinical evidence on pioglitazone pharmacogenomics. *Frontiers in Pharmacology*, *4*. https://doi.org/10.3389/fphar.2013.00147

Kazmierski, J., Messini-Zachou, C., Gkioka, M., & Tsolaki, M. (2018). The impact of a long-term rivastigmine and donepezil treatment on all-cause mortality in patients with Alzheimer's disease. *American Journal of Alzheimer's Disease & Other Dementias*, *33*(6), 385–393. https://doi.org/10.1177/1533317518775044

Khan, A. H., Zou, Z., Xiang, Y., Chen, S., & Tian, X.-L. (2019). Conserved signaling pathways genetically associated with longevity across the species. *Biochimica et Biophysica Acta (BBA) - Molecular Basis of Disease*, *1865*(7), 1745–1755. https://doi.org/10.1016/j.bbadis.2018.09.001

Khomtchouk, B. B., Sun, P., Ditmarsch, M., Kastelein, J. J. P., & Davidson, M. H. (2023). *CETP and SGLT2 inhibitor combination therapy improves glycemic control*. Cold Spring Harbor Laboratory. http://dx.doi.org/10.1101/2023.06.13.23291357

Khoury, R., Rajamanickam, J., & Grossberg, G. T. (2018). An update on the safety of current therapies for Alzheimer's disease: Focus on rivastigmine. *Therapeutic Advances in Drug Safety*, *9*(3), 171–178. https://doi.org/10.1177/2042098617750555

Kim, J., Ahn, C. W., Fang, S., Lee, H. S., & Park, J. S. (2019). Association between metformin dose and vitamin B12 deficiency in patients with type 2 diabetes. *Medicine*, *98*(46), e17918. https://doi.org/10.1097/md.0000000000017918

Kim, S.-H., Sanak, M., & Park, H.-S. (2013). Genetics of hypersensitivity to aspirin and nonsteroidal anti-inflammatory drugs. *Immunology and Allergy Clinics of North America*, *33*(2), 177–194. https://doi.org/10.1016/j.iac.2012.10.003

Lambert, S. A., Gil, L., Jupp, S., Ritchie, S. C., Xu, Y., Buniello, A., McMahon, A., Abraham, G., Chapman, M., Parkinson, H., Danesh, J., MacArthur, J. A. L., & Inouye, M. (2021). The Polygenic Score Catalog as an open database for reproducibility and systematic evaluation. *Nature Genetics*, *53*(4), 420–425. https://doi.org/10.1038/s41588-021-00783-5

Landrum, M. J., Lee, J. M., Benson, M., Brown, G. R., Chao, C., Chitipiralla, S., Gu, B., Hart, J., Hoffman, D., Jang, W., Karapetyan, K., Katz, K., Liu, C., Maddipatla, Z., Malheiro, A., McDaniel, K., Ovetsky, M., Riley, G., Zhou, G., … Maglott, D. R. (2017). ClinVar: Improving access to variant interpretations and supporting evidence. *Nucleic Acids Research*, *46*(D1), D1062–D1067. https://doi.org/10.1093/nar/gkx1153

Lawson, D. J., van Dorp, L., & Falush, D. (2018). A tutorial on how not to over-interpret STRUCTURE and ADMIXTURE bar plots. *Nature Communications*, *9*(1). https://doi.org/10.1038/s41467-018-05257-7

Ledford, H. (2012). Sirtuin protein linked to longevity in mammals. *Nature*. https://doi.org/10.1038/nature.2012.10074

Li, X., Li, C., Zhang, W., Wang, Y., Qian, P., & Huang, H. (2023). Inflammation and aging: Signaling pathways and intervention therapies. *Signal Transduction and Targeted Therapy*, *8*(1). https://doi.org/10.1038/s41392-023-01502-8

Li, Y., Byun, J., Cai, G., Xiao, X., Han, Y., Cornelis, O., Dinulos, J. E., Dennis, J., Easton, D., Gorlov, I., Seldin, M. F., & Amos, C. I. (2016). FastPop: A rapid principal component derived method to infer intercontinental ancestry using genetic data. *BMC Bioinformatics*, *17*(1). https://doi.org/10.1186/s12859-016-0965-1

Lin, J.-R., Sin-Chan, P., Napolioni, V., Torres, G. G., Mitra, J., Zhang, Q., Jabalameli, M. R., Wang, Z., Nguyen, N., Gao, T., Laudes, M., Görg, S., Franke, A., Nebel, A., Greicius, M. D., Atzmon, G., Ye, K., Gorbunova, V., Ladiges, W. C., … Zhang, Z. D. (2021). Rare genetic coding variants associated with human longevity and protection against age-related diseases. *Nature Aging*, *1*(9), 783–794.

https://doi.org/10.1038/s43587-021-00108-5

Łukasik, P., Załuski, M., & Gutowska, I. (2021). Cyclin-Dependent kinases (CDK) and their role in diseases development–review. *International Journal of Molecular Sciences*, *22*(6), 2935. https://doi.org/10.3390/ijms22062935

Lunetta, K. L., D'Agostino, R. B., Sr, Karasik, D., Benjamin, E. J., Guo, C.-Y., Govindaraju, R., Kiel, D. P., Kelly-Hayes, M., Massaro, J. M., Pencina, M. J., Seshadri, S., & Murabito, J. M. (2007). Genetic correlates of longevity and selected age-related phenotypes: A genome-wide association study in the Framingham Study. *BMC Medical Genetics*, *8*(S1). https://doi.org/10.1186/1471-2350-8-s1-s13

Mannick, J. B., & Lamming, D. W. (2023). Targeting the biology of aging with mTOR inhibitors. *Nature Aging*, *3*(6), 642–660. https://doi.org/10.1038/s43587-023-00416-y

Matey-Hernandez, M. L., Williams, F. M. K., Potter, T., Valdes, A. M., Spector, T. D., & Menni, C. (2018). Genetic and microbiome influence on lipid metabolism and dyslipidemia. *Physiological Genomics*, *50*(2), 117–126. https://doi.org/10.1152/physiolgenomics.00053.2017

Melzer, D., Pilling, L. C., & Ferrucci, L. (2019). The genetics of human ageing. *Nature Reviews Genetics*, *21*(2), 88–101. https://doi.org/10.1038/s41576-019-0183-6

Min, Z., Mi, Y., Lv, Z., Sun, Y., Tang, B., Wu, H., Zhang, Z., Pan, H., Zhang, Y., Lu, C., Zuo, L., & Zhang, L. (2022). Associations of Genetic Polymorphisms of mTOR rs2295080 T/G and rs1883965 G/A with Susceptibility of Urinary System Cancers. *Disease Markers*, *2022*, 1–16. https://doi.org/10.1155/2022/1720851

Mogi, M. (2020). Effect of renin–angiotensin system on senescence. *Geriatrics & Gerontology International*, *20*(6), 520–525. https://doi.org/10.1111/ggi.13927

Mohammed, I., Hollenberg, M. D., Ding, H., & Triggle, C. R. (2021). A critical review of the evidence that Metformin is a putative anti-aging drug that enhances healthspan and extends lifespan. *Frontiers in Endocrinology*, *12*. https://doi.org/10.3389/fendo.2021.718942

Moskalev, A., Guvatova, Z., Lopes, I. D. A., Beckett, C. W., Kennedy, B. K., De Magalhaes, J. P., & Makarov, A. A. (2022). Targeting aging mechanisms: Pharmacological perspectives. *Trends in Endocrinology & Metabolism*, *33*(4), 266–280. https://doi.org/10.1016/j.tem.2022.01.007

Mutlu, A. S., Duffy, J., & Wang, M. C. (2021). Lipid metabolism and lipid signals in aging and longevity. *Developmental Cell*, *56*(10), 1394–1407. https://doi.org/10.1016/j.devcel.2021.03.034

Nasykhova, Y., Barbitoff, Y., Tonyan, Z., Danilova, M., Nevzorov, I., Komandresova, T., Mikhailova, A.,

Vasilieva, T., Glavnova, O., Yarmolinskaya, M., Sluchanko, E., & Glotov, A. (2022). Genetic and phenotypic factors affecting glycemic response to metformin therapy in patients with type 2 diabetes mellitus. *Genes*, *13*(8), 1310. https://doi.org/10.3390/genes13081310

Navar-Boggan, A. M., Peterson, E. D., D'Agostino, R. B., Sr, Neely, B., Sniderman, A. D., & Pencina, M. J. (2015). Hyperlipidemia in early adulthood increases long-term risk of coronary heart disease. *Circulation*, *131*(5), 451–458. https://doi.org/10.1161/circulationaha.114.012477

Newcombe, E. A., Camats-Perna, J., Silva, M. L., Valmas, N., Huat, T. J., & Medeiros, R. (2018). Inflammation: The link between comorbidities, genetics, and Alzheimer's disease. *Journal of Neuroinflammation*, *15*(1). https://doi.org/10.1186/s12974-018-1313-3

Oestereich, F., Yousefpour, N., Yang, E., Phénix, J., Nezhad, Z. S., Nitu, A., Vázquez Cobá, A., Ribeiro-da-Silva, A., Chaurand, P., & Munter, L. M. (2022). The cholesteryl ester transfer protein (CETP) raises cholesterol levels in the brain. *Journal of Lipid Research*, *63*(9), 100260. https://doi.org/10.1016/j.jlr.2022.100260

Palaniswamy, C., Sekhri, A., Aronow, W. S., Kalra, A., & Peterson, S. J. (2011). Association of warfarin use with valvular and vascular calcification: A review. *Clinical Cardiology*, *34*(2), 74–81. https://doi.org/10.1002/clc.20865

Papadopoli, D., Boulay, K., Kazak, L., Pollak, M., Mallette, F., Topisirovic, I., & Hulea, L. (2019). mTOR as a central regulator of lifespan and aging. *F1000Research*, *8*, 998. https://doi.org/10.12688/f1000research.17196.1

Passarino, G., De Rango, F., & Montesanto, A. (2016). Human longevity: Genetics or Lifestyle? It takes two to tango. *Immunity & Ageing*, *13*(1). https://doi.org/10.1186/s12979-016-0066-z

Passtoors, W. M., Beekman, M., Deelen, J., van der Breggen, R., Maier, A. B., Guigas, B., Derhovanessian, E., van Heemst, D., de Craen, A. J. M., Gunn, D. A., Pawelec, G., & Slagboom, P. E. (2012). Gene expression analysis of mTOR pathway: Association with human longevity. *Aging Cell*, *12*(1), 24–31. https://doi.org/10.1111/acel.12015

Pastori, D., Cormaci, V. M., Marucci, S., Franchino, G., Del Sole, F., Capozza, A., Fallarino, A., Corso, C., Valeriani, E., Menichelli, D., & Pignatelli, P. (2023). A Comprehensive Review of Risk Factors for Venous Thromboembolism: From Epidemiology to Pathophysiology. *International Journal of Molecular Sciences*, *24*(4), 3169. https://doi.org/10.3390/ijms24043169

Peinado-Ruiz, I. C., Burgos-Molina, A. M., Sendra-Portero, F., & Ruiz-Gómez, M. J. (2022). Relationship

between heat shock proteins and cellular resistance to drugs and ageing. *Experimental Gerontology*, *167*, 111896. https://doi.org/10.1016/j.exger.2022.111896

Perna, S., Guido, D., Bologna, C., Solerte, S. B., Guerriero, F., Isu, A., & Rondanelli, M. (2016). Liraglutide and obesity in elderly: Efficacy in fat loss and safety in order to prevent sarcopenia. A perspective case series study. *Aging Clinical and Experimental Research*, *28*(6), 1251–1257. https://doi.org/10.1007/s40520-015-0525-y

Pilling, L. C., Atkins, J. L., Bowman, K., Jones, S. E., Tyrrell, J., Beaumont, R. N., Ruth, K. S., Tuke, M. A., Yaghootkar, H., Wood, A. R., Freathy, R. M., Murray, A., Weedon, M. N., Xue, L., Lunetta, K., Murabito, J. M., Harries, L. W., Robine, J.-M., Brayne, C., … Melzer, D. (2016). Human longevity is influenced by many genetic variants: Evidence from 75,000 UK Biobank participants. *Aging*, *8*(3), 547–560. https://doi.org/10.18632/aging.100930

Pilling, L. C., Kuo, C.-L., Sicinski, K., Tamosauskaite, J., Kuchel, G. A., Harries, L. W., Herd, P., Wallace, R., Ferrucci, L., & Melzer, D. (2017). Human longevity: 25 genetic loci associated in 389,166 UK biobank participants. *Aging*, *9*(12), 2504–2520. https://doi.org/10.18632/aging.101334

Rahimi, Z. (2016). The role of renin angiotensin aldosterone system genes in diabetic nephropathy. *Canadian Journal of Diabetes*, *40*(2), 178–183. https://doi.org/10.1016/j.jcjd.2015.08.016

Rajagopalan, R., Perez, A., Ye, Z., Khan, M., & Murray, F. T. (2004). Pioglitazone is Effective Therapy for Elderly Patients with Type 2 Diabetes Mellitus. *Drugs & Aging*, *21*(4), 259–271. https://doi.org/10.2165/00002512-200421040-00004

Ray, B., Maloney, B., Sambamurti, K., Karnati, H. K., Nelson, P. T., Greig, N. H., & Lahiri, D. K. (2020). Rivastigmine modifies the α-secretase pathway and potentially early Alzheimer's disease. *Translational Psychiatry*, *10*(1). https://doi.org/10.1038/s41398-020-0709-x

Ray, K. K., Ference, B. A., Séverin, T., Blom, D., Nicholls, S. J., Shiba, M. H., Almahmeed, W., Alonso, R., Daccord, M., Ezhov, M., Olmo, R. F., Jankowski, P., Lanas, F., Mehta, R., Puri, R., Wong, N. D., Wood, D., Zhao, D., Gidding, S. S., … Santos, R. D. (2022). World heart federation cholesterol roadmap 2022. *Global Heart*, *17*(1), 75. https://doi.org/10.5334/gh.1154

Revelas, M., Thalamuthu, A., Oldmeadow, C., Evans, T.-J., Armstrong, N. J., Kwok, J. B., Brodaty, H., Schofield, P. R., Scott, R. J., Sachdev, P. S., Attia, J. R., & Mather, K. A. (2018). Review and meta-analysis of genetic polymorphisms associated with exceptional human longevity. *Mechanisms of Ageing and Development*, *175*, 24–34. https://doi.org/10.1016/j.mad.2018.06.002

rkimoakbioinformatics. (n.d.). *GitHub - Rkimoakbioinformatics/oakvar: Genomic variant analysis platform*. GitHub. Retrieved August 18, 2023, from https://github.com/rkimoakbioinformatics/oakvar

Rosada, A., Kassner, U., Weidemann, F., König, M., Buchmann, N., Steinhagen-Thiessen, E., & Spira, D. (2020). Hyperlipidemias in elderly patients: Results from the Berlin Aging Study II (BASEII), a cross-sectional study. *Lipids in Health and Disease*, *19*(1). https://doi.org/10.1186/s12944-020-01277-9

Rudic, B., Schimpf, R., & Borggrefe, M. (2014). Short QT syndrome – review of diagnosis and treatment. *Arrhythmia & Electrophysiology Review*, *3*(2), 76. https://doi.org/10.15420/aer.2014.3.2.76

Sampson, J. N., Kidd, K. K., Kidd, J. R., & Zhao, H. (2011). Selecting snps to identify ancestry. *Annals of Human Genetics*, *75*(4), 539–553. https://doi.org/10.1111/j.1469-1809.2011.00656.x

Santos, A. L., & Lindner, A. B. (2017). Protein posttranslational modifications: Roles in aging and age-related disease. *Oxidative Medicine and Cellular Longevity*, *2017*, 1–19. https://doi.org/10.1155/2017/5716409

Sebastiani, P., Gurinovich, A., Nygaard, M., Sasaki, T., Sweigart, B., Bae, H., Andersen, S. L., Villa, F., Atzmon, G., Christensen, K., Arai, Y., Barzilai, N., Puca, A., Christiansen, L., Hirose, N., & Perls, T. T. (2018). APOEAlleles and extreme human longevity. *The Journals of Gerontology: Series A*, *74*(1), 44–51. https://doi.org/10.1093/gerona/gly174

Sebastiani, P., & Perls, T. T. (2012). The Genetics of Extreme Longevity: Lessons from the New England centenarian study. *Frontiers in Genetics*, *3*. https://doi.org/10.3389/fgene.2012.00277

Sebastiani, P., Solovieff, N., DeWan, A. T., Walsh, K. M., Puca, A., Hartley, S. W., Melista, E., Andersen, S., Dworkis, D. A., Wilk, J. B., Myers, R. H., Steinberg, M. H., Montano, M., Baldwin, C. T., Hoh, J., & Perls, T. T. (2012). Genetic signatures of exceptional longevity in humans. *PLoS ONE*, *7*(1), e29848. https://doi.org/10.1371/journal.pone.0029848

Seo, Y.-G. (2021). Side effects associated with liraglutide treatment for obesity as well as diabetes. *Journal of Obesity & Metabolic Syndrome*, *30*(1), 12–19. https://doi.org/10.7570/jomes20059

Serrano-Pozo, A., Das, S., & Hyman, B. T. (2021). APOE and Alzheimer's disease: Advances in genetics, pathophysiology, and therapeutic approaches. *The Lancet Neurology*, *20*(1), 68–80. https://doi.org/10.1016/s1474-4422(20)30412-9

Shamanskiy, V., Mikhailova, A. A., Ushakova, K., Mikhailova, A. G., Oreshkov, S., Knorre, D., Tretiakov, E. O., Ri, N., Overdevest, J. B., Lukowski, S. W., Gostimskaya, I., Yurov, V., Liou, C.-W., Lin, T.-K., Kunz, W. S., Reymond, A., Mazunin, I., Bazykin, G. A., Gunbin, K., … Popadin, K. (2019). *Risk of mitochondrial deletions is affected by the global secondary structure of the human mitochondrial genome*. Cold Spring

Harbor Laboratory. http://dx.doi.org/10.1101/603282

Shen, D., Li, H., Zhou, R., Liu, M., Yu, H., & Wu, D.-F. (2018). Pioglitazone attenuates aging-related disorders in aged apolipoprotein E deficient mice. *Experimental Gerontology*, *102*, 101–108. https://doi.org/10.1016/j.exger.2017.12.002

Soerensen, M., Dato, S., Tan, Q., Thinggaard, M., Kleindorp, R., Beekman, M., Suchiman, H. E. D., Jacobsen, R., McGue, M., Stevnsner, T., Bohr, V. A., de Craen, A. J. M., Westendorp, R. G. J., Schreiber, S., Slagboom, P. E., Nebel, A., Vaupel, J. W., Christensen, K., & Christiansen, L. (2012). Evidence from case–control and longitudinal studies supports associations of genetic variation in APOE, CETP, and IL6 with human longevity. *AGE*, *35*(2), 487–500. https://doi.org/10.1007/s11357-011-9373-7

Stelzer, G., Rosen, N., Plaschkes, I., Zimmerman, S., Twik, M., Fishilevich, S., Stein, T. I., Nudel, R., Lieder, I., Mazor, Y., Kaplan, S., Dahary, D., Warshawsky, D., Guan‐Golan, Y., Kohn, A., Rappaport, N., Safran, M., & Lancet, D. (2016). The GeneCards Suite: From gene data mining to disease genome sequence analyses. *Current Protocols in Bioinformatics*, *54*(1). https://doi.org/10.1002/cpbi.5

Tacutu, R., Thornton, D., Johnson, E., Budovsky, A., Barardo, D., Craig, T., Diana, E., Lehmann, G., Toren, D., Wang, J., Fraifeld, V. E., & de Magalhães, J. P. (2017). Human Ageing Genomic Resources: New and updated databases. *Nucleic Acids Research*, *46*(D1), D1083–D1090. https://doi.org/10.1093/nar/gkx1042

Talukder, A., Barham, C., Li, X., & Hu, H. (2020). Interpretation of deep learning in genomics and epigenomics. *Briefings in Bioinformatics*, *22*(3). https://doi.org/10.1093/bib/bbaa177

Tesi, N., van der Lee, S. J., Hulsman, M., Jansen, I. E., Stringa, N., van Schoor, N. M., Scheltens, P., van der Flier, W. M., Huisman, M., Reinders, M. J. T., & Holstege, H. (2020). Polygenic risk score of longevity predicts longer survival across an age continuum. *The Journals of Gerontology: Series A*, *76*(5), 750–759. https://doi.org/10.1093/gerona/glaa289

Toma, T. T., Dawson, J. M., & Adjeroh, D. A. (2018). Human ancestry indentification under resource constraints -- what can one chromosome tell us about human biogeographical ancestry? *BMC Medical Genomics*, *11*(S5). https://doi.org/10.1186/s12920-018-0412-4

Tonoike, M., Chujo, D., & Noda, M. (2018). Efficacy and safety of liraglutide added to insulin therapy in elderly patients with type 2 diabetes. *Endocrinology, Diabetes & Metabolism*, *2*(1). https://doi.org/10.1002/edm2.43

Ukraintseva, S., Duan, M., Arbeev, K., Wu, D., Bagley, O., Yashkin, A. P., Gorbunova, G., Akushevich, I.,

Kulminski, A., & Yashin, A. (2021). Interactions between genes from aging pathways may influence human lifespan and improve animal to human translation. *Frontiers in Cell and Developmental Biology*, *9*(692020). https://doi.org/10.3389/fcell.2021.692020

van den Berg, N., Rodríguez-Girondo, M., van Dijk, I. K., Mourits, R. J., Mandemakers, K., Janssens, A. A. P. O., Beekman, M., Smith, K. R., & Slagboom, P. E. (2019). Longevity defined as top 10% survivors and beyond is transmitted as a quantitative genetic trait. *Nature Communications*, *10*(1). https://doi.org/10.1038/s41467-018-07925-0

van Hilten, A., Kushner, S. A., Kayser, M., Ikram, M. A., Adams, H. H. H., Klaver, C. C. W., Niessen, W. J., & Roshchupkin, G. V. (2021). GenNet framework: Interpretable deep learning for predicting phenotypes from genetic data. *Communications Biology*, *4*(1). https://doi.org/10.1038/s42003-021-02622-z

Vargas-Soria, M., Carranza-Naval, M. J., del Marco, A., & Garcia-Alloza, M. (2021). Role of liraglutide in Alzheimer's disease pathology. *Alzheimer's Research & Therapy*, *13*(1). https://doi.org/10.1186/s13195-021-00853-0

Vitale, G., Barbieri, M., Kamenetskaya, M., & Paolisso, G. (2017). GH/IGF-I/insulin system in centenarians. *Mechanisms of Ageing and Development*, *165*(Pt B), 107–114. https://doi.org/10.1016/j.mad.2016.12.001

Vitale, G., Pellegrino, G., Vollery, M., & Hofland, L. J. (2019). ROLE of IGF-1 system in the modulation of longevity: Controversies and new insights from a centenarians' perspective. *Frontiers in Endocrinology*, *10*. https://doi.org/10.3389/fendo.2019.00027

Wang, C.-H., Li, F., & Takahashi, N. (2010). The renin angiotensin system and the metabolic syndrome. *Open Hypertens J*, *3*(1–13). https://doi.org/10.2174/1876526203010001

Watroba, M., & Szukiewicz, D. (2021). Sirtuins at the service of healthy longevity. *Frontiers in Physiology*, *12*. https://doi.org/10.3389/fphys.2021.724506

Whirl‐Carrillo, M., Huddart, R., Gong, L., Sangkuhl, K., Thorn, C. F., Whaley, R., & Klein, T. E. (2021). An evidence‐based framework for evaluating pharmacogenomics knowledge for personalized medicine. *Clinical Pharmacology & Therapeutics*, *110*(3), 563–572. https://doi.org/10.1002/cpt.2350

Whitley, K. V., Tueller, J. A., & Weber, K. S. (2020). Genomics education in the era of personal genomics: Academic, professional, and public considerations. *International Journal of Molecular Sciences*, *21*(3), 768. https://doi.org/10.3390/ijms21030768

Wierman, M. B., & Smith, J. S. (2013). Yeast sirtuins and the regulation of aging. *FEMS Yeast Research*, *14*(1),

73–88. https://doi.org/10.1111/1567-1364.12115

Wingfield, B., Lambert, S., & Gil, L. (2023, December 5). *PGScatalog/pgsc_calc: pgsc_calc v2.0.0-alpha.4*. Zenodo. https://zenodo.org/doi/10.5281/zenodo.6617030

Wu, H.-X., Liu, K.-K., Li, B.-N., Liu, S., & Jin, J.-C. (2022). Efficacy and safety of sacubitril/valsartan in the treatment of middle-aged and elderly patients with hypertension: A systematic review and meta-analysis of randomized controlled trials. *Annals of Palliative Medicine*, *11*(5), 1811–1825. https://doi.org/10.21037/apm-22-503

Xu, L., Ma, X., Verma, N., Perie, L., Pendse, J., Shamloo, S., Marie Josephson, A., Wang, D., Qiu, J., Guo, M., Ping, X., Allen, M., Noguchi, A., Springer, D., Shen, F., Liu, C., Zhang, S., Li, L., Li, J., … Mueller, E. (2020). PPARγ agonists delay age‐associated metabolic disease and extend longevity. *Aging Cell*, *19*(11). https://doi.org/10.1111/acel.13267

Yang, A., Kantor, B., & Chiba-Falek, O. (2021). APOE: The new frontier in the development of a therapeutic target towards precision medicine in late-onset alzheimer's. *International Journal of Molecular Sciences*, *22*(3), 1244. https://doi.org/10.3390/ijms22031244

Yashin, A. I., Wu, D., Arbeev, K. G., & Ukraintseva, S. V. (2010). Joint influence of small-effect genetic variants on human longevity. *Aging*, *2*(9), 612–620. https://doi.org/10.18632/aging.100191

Yashin, A. I., Wu, D., Arbeeva, L. S., Arbeev, K. G., Kulminski, A. M., Akushevich, I., Kovtun, M., Culminskaya, I., Stallard, E., Li, M., & Ukraintseva, S. V. (2015). Genetics of aging, health, and survival: Dynamic regulation of human longevity related traits. *Frontiers in Genetics*, *6*. https://doi.org/10.3389/fgene.2015.00122

Yates, S. C., Zafar, A., Hubbard, P., Nagy, S., Durant, S., Bicknell, R., Wilcock, G., Christie, S., Esiri, M. M., Smith, A. D., & Nagy, Z. (2013). Dysfunction of the mTOR pathway is a risk factor for Alzheimer's disease. *Acta Neuropathologica Communications*, *1*(1). https://doi.org/10.1186/2051-5960-1-3

Zahid, A., Li, B., Kombe, A. J. K., Jin, T., & Tao, J. (2019). Pharmacological inhibitors of the NLRP3 inflammasome. *Frontiers in Immunology*, *10*. https://doi.org/10.3389/fimmu.2019.02538

Zamani, M., Mohammadi, M., Zamani, H., & Tavasoli, A. (2015). Pharmacogenetic study on the impact of rivastigmine concerning genetic variants of A2M and IL-6 genes on iranian alzheimer's patients. *Molecular Neurobiology*, *53*(7), 4521–4528. https://doi.org/10.1007/s12035-015-9387-8

Zdravkovic, S., Wienke, A., Pedersen, N. L., Marenberg, M. E., Yashin, A. I., & De Faire, U. (2002). Heritability of death from coronary heart disease: A 36-year follow-up of 20 966 Swedish twins. *Journal of Internal*

*Medicine*, *252*(3), 247–254. https://doi.org/10.1046/j.1365-2796.2002.01029.x

Zeng, Y., Nie, C., Min, J., Liu, X., Li, M., Chen, H., Xu, H., Wang, M., Ni, T., Li, Y., Yan, H., Zhang, J.-P., Song, C., Chi, L.-Q., Wang, H.-M., Dong, J., Zheng, G.-Y., Lin, L., Qian, F., … Vaupel, J. W. (2016). Novel loci and pathways significantly associated with longevity. *Scientific Reports*, *6*(1). https://doi.org/10.1038/srep21243

Zhang, H., Chu, Y., Zheng, H., Wang, J., Song, B., & Sun, Y. (2020). Liraglutide improved the cognitive function of diabetic mice via the receptor of advanced glycation end products down-regulation. *Aging*, *13*(1), 525–536. https://doi.org/10.18632/aging.202162

Zhang, M., Yan, W., Yu, Y., Cheng, J., Yi, X., Guo, T., Liu, N., Shang, J., Wang, Z., Hu, H., & Chen, L. (2021). Liraglutide ameliorates diabetes-associated cognitive dysfunction via rescuing autophagic flux. *Journal of Pharmacological Sciences*, *147*(3), 234–244. https://doi.org/10.1016/j.jphs.2021.07.004

Zhang, Z. D., Milman, S., Lin, J.-R., Wierbowski, S., Yu, H., Barzilai, N., Gorbunova, V., Ladiges, W. C., Niedernhofer, L. J., Suh, Y., Robbins, P. D., & Vijg, J. (2020). Genetics of extreme human longevity to guide drug discovery for healthy ageing. *Nature Metabolism*, *2*(8), 663–672. https://doi.org/10.1038/s42255-020-0247-0

Zhao, L., Cao, J., Hu, K., He, X., Yun, D., Tong, T., & Han, L. (2020). Sirtuins and their Biological Relevance in Aging and Age-Related Diseases. *Aging and Disease*, *11*(4), 927. https://doi.org/10.14336/ad.2019.0820

Zhu, X., Chang, Y.-P. C., Yan, D., Weder, A., Cooper, R., Luke, A., Kan, D., & Chakravarti, A. (2003). Associations between hypertension and genes in the renin-angiotensin system. *Hypertension*, *41*(5), 1027–1034. https://doi.org/10.1161/01.hyp.0000068681.69874.cb